\documentclass[a4paper,amsfonts, amssymb, amsmath, reprint, showkeys,twoside,onecolumn,pra]{revtex4-1}
\usepackage[english]{babel}
\usepackage[utf8]{inputenc}
\usepackage{xcolor}
\usepackage{color}
\usepackage{amssymb}
\usepackage[colorinlistoftodos, color=green!40, prependcaption]{todonotes}
\usepackage[pdftex, pdftitle={Article}, pdfauthor={Author}]{hyperref} 
\usepackage{mhchem}
\usepackage{physics}

\begin{document}
\title{Anisotropic Heisenberg Su-Schrieffer-Heeger spin chain as a quantum channel}

\author{Lautaro Moragues$^{(1)}$}
\author{Diego Acosta Coden$^{(1)}$}
\author{Omar Osenda$^{(2)}$}
\author{Alejandro Ferr\'on$^{(1)}$}

    \email[Correspondence email address: ]{aferron@exa.unne.edu.ar}
    \affiliation{${(1)}$ Instituto de Modelado e Innovación Tecnológica (CONICET-UNNE) and Facultad de Ciencias Exactas, Naturales y Agrimensura, Universidad Nacional del Nordeste, Avenida Libertad 5400, W3404AAS Corrientes, Argentina.\\
    ${(2)}$ Instituto de Física Enrique Gaviola (CONICET-UNC) and Facultad de Matemática, Astronomía, Física y Computación, Universidad Nacional de Córdoba, Av. Medina Allende s/n, Ciudad Universitaria, CP:X5000HUA Córdoba, Argentina.}

\date{\today} 

\begin{abstract}
Quantum state transmission in spin chains is a fundamental problem within quantum technologies. The Su-Schrieffer-Heeger (SSH) model, first introduced in the context of polyacetylene, provides a paradigmatic example of a system exhibiting topological and non-topological phases. We explore the transmission of one and two excitations in anisotropic Heisenberg SSH spin chains and analyze the relationship between topological properties and state transfer efficiency. We examine the robustness of quantum state transmission against static disorder in the trivial and topological regimes, exploring how topological protection influences transmission fidelity. We also consider the effect of dipolar interactions, introducing long-range couplings and breaking the conservation of total magnetization. Furthermore, we employ optimal control theory to design driving pulses for state transmission, finding substantial differences between optimizing in the trivial and topological regimes. Our results provide insights about the interplay between topology, disorder, interactions, and control strategies in quantum state transfer.

\end{abstract}


\maketitle

\section{Introduction}

Efficient quantum state transmission \cite{Xie2023,Maleki2021} is a key point for developing scalable quantum technologies, enabling the reliable transfer of information across quantum networks. Spin chains are essential theoretical models that allow the study of different possible experimental platforms, such as transmon qubit chains \cite{Stehlik2021,serra2025perfect,Li2018transmon} or arrays of magnetic atoms. In this context, numerous studies focus on topological spin chains due to their distinctive properties, amongst them the presence of localized edge states that are robust against perturbations and their spectral properties. The Su-Schrieffer-Heeger (SSH) model, originally introduced to describe polyacetylene \cite{Su1979}, has become a paradigmatic system in this field, as it hosts topologically protected edge states for some value of the parameters that define the system \cite{Xie2019}. These states are resilient against certain types of perturbations, such as static disorder \cite{Estarellas2017}. This topological protection has made SSH-like models highly relevant for quantum information processing \cite{Wang2022}, motivating their study as quantum channels with enhanced robustness against disorder and imperfections.

Despite the promised protection properties,  the possible use of topological states in quantum information protocols is far from given \cite{Qi2024,legg2025commentinasalhybriddevices,Aghaee2023hybridtopo}. There are numerous problems with the actual preparation of topological states, their transmission, and manipulation. In topological chains, the energy of the topological states lies in the middle of the gap between bands of states. These circumstances render dissipative preparation methods almost useless. This picture differs from the scenario in topological superconductors, where the topological state has less energy than the normal states. 

Regardless of the controllability of quantum chains \cite{Wang2016,Burgarth2010,Yang2010,Gong2007,Murphy2010},  the controlled transition from a given reference state to a topological one is difficult and time-consuming because its energy lies within the energy gap. So, what protects the energy of a topological state is what makes it almost inaccessible through the usual navigation of the spectrum landscape. 

Before using a topological chain as a transmission channel, it is necessary to know where we want to encode the information. To employ the protection, we should encode the information in a topological state. As the topological state is an eigenstate, its time evolution is trivial unless we globally change the Hamiltonian, from one with a topological state localized at one extreme of the chain to another with its topological state localized at the other extreme of the chain. Another possibility relies on a Hamiltonian with topological states located at both ends of the chains \cite{Qi2024}. In this case, an adequate linear combination produces the switching of the superposition state between the extremes of the chain. Both possibilities bring an extremely slow state transfer. It is worth asking about the necessity or even the convenience of coding the information in a topological eigenstate. 

Several proposals have explored the implementation of topological spin chains using different experimental platforms. Recent studies considered chains with topological states as channels for quantum state transfer \cite{Wang2022,serra2025perfect,Maleki2021,Lemonde2019,Romero2024}, but it remains unclear whether such systems are optimal for quantum information transmission. Wang and coworkers \cite{wang2024TopoMagneticConstruction} built dimerized chains of titanium atoms deposited on MgO/Ag surfaces, using a scanning tunneling microscope (STM), individual Ti atoms can be placed on specific sites, either atop oxygen atoms or at bridge positions, allowing precise control over the interatomic spacing and enabling the formation of dimerized configurations. Moreover, in this work, the topological properties of the resulting spin chains were experimentally probed using STM-based electron spin resonance (ESR-STM) techniques \cite{baumann2015b}. More recently,  through on-surface synthesis, alternating-exchange Heisenberg spin $1/2$ chains have been constructed using graphene-based nanostructures that host localized spin $1/2$ moments \cite{zhao2024nanografeno1,zhao2025nanografeno2,fu2025nanografenobuilding}. In this case, the authors employed STM for the characterization of spin excitations by inelastic electron tunneling spectroscopy. 

We aim to test the ability of chains with topological states to transfer one and two-qubit arbitrary states initially prepared on one extreme of the chain. While coding the information of a one-qubit arbitrary state on a topological state seems appealing because a one-excitation state is as strongly localized as the topological state is, this is not true for arbitrary two-qubit states, whose localization could be equally distributed over the first two qubits at the extreme of the chain, or even bigger in the second qubit than in the first. This, without mentioning the difficulties associated with retrieving information of a topological state spread over several qubits. We will consider a chain Hamiltonian equivalent to a dimerized anisotropic Heisenberg chain with only nearest-neighbors interactions. Additionally, we will study the effect of a dipolar long-range term on the transfer properties. This last term plays the role of unwanted interactions that are present in the system and worsen transmission. For instance, in magnetic atoms located over non-magnetic surfaces, the dipolar term comes from the dipolar interaction between atoms, while the Heisenberg term comes from the exchange interaction. The dimerization of the anisotropic Heisenberg part is responsible for the appearance of topological states.

We organized the paper as follows. In Sec. \ref{smc}, we introduce the spin XXZ-SSH Hamiltonian, with and without dipolar interactions. We also describe how we implement the analysis of the disorder, how we compute state transfer probabilities and fidelities for one and two excitations, and how we apply optimal control theory to design efficient transmission protocols. In Sec. \ref{srat1e}, we study the autonomous transmission of a single excitation, paying particular attention to the efficiency for different parameters and analyzing the effect of static disorder. In Sec. \ref{srat2e}, we perform the same analysis for the case of transmitting two excitations. In Sec. \ref{srdip}, we examine the effect of dipolar interactions, discussing the challenges and potential benefits they present. Sec. \ref{sroct} is devoted to the application of optimal control theory in designing high-fidelity transmission protocols, and we examine how the controllability of the chain depends on the  Hamiltonian parameters. Throughout all sections, we contrast the behavior in topological and non-topological regimes. Finally, Sec. \ref{sc} presents our main conclusions.


\section{Model Hamiltonian, Excitation Transmission and Calculation Details}\label{smc}

In this Section, we present the chain Hamiltonian responsible for the autonomous evolution of the system, the quantities that quantify the quality of the transmission protocol for arbitrary one and two-qubit quantum state transfer, and the averaged fidelities that we will use later on to study the properties of the transmission protocol when some static disorder may appers on the chains. Additionally, to attain faster and better quantum state transfer than is possible with autonomous evolution, we briefly discuss the mathematical details necessary to implement Optimal Control Theory.

\subsection{Model Hamiltonian}
\label{smh}

\begin{figure}[h]
 \centering
 \includegraphics[width=0.6\textwidth]{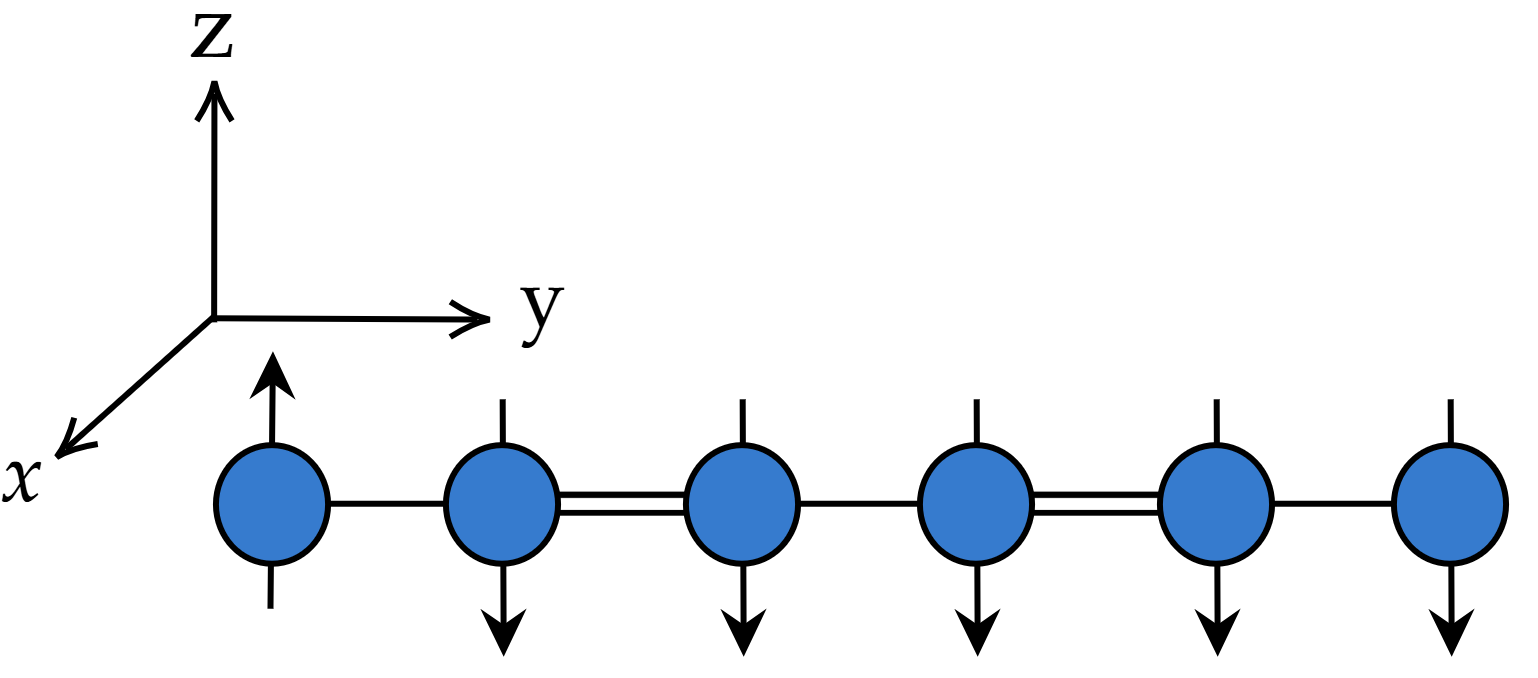}
 \caption{Scheme of a dimerized spin chain.}
 \label{fig1}
\end{figure}

The Hamiltonian for a ferromagnetic SSH-XXZ spin chain with equally spaced sites, as the one depicted in Fig. \ref{fig1}, can be written as

\begin{equation}\label{eqhssh}
\hat{H}_{ssh}=-\frac{1}{4}\sum_{i=1}^{N-1}J_i\left[\sigma_j^x\sigma_{i+1}^x+\sigma_i^y\sigma_{i+1}^y+\Delta\sigma_i^z\sigma_{j+1}^z\right],
\end{equation}

\noindent where $J_i=J\left(1+(-1)^i\,\eta\right)$, $N$ is the total number of spins, $\eta$ is the dimerization parameter, $\Delta$ is the perpendicular anisotropy, and $\sigma_j^{a}$ with $a=x,y,z$ are the Pauli matrices.

Throughout this paper, we analyze the dynamics of this Hamiltonian as a function of its parameters $\eta$ and $\Delta$. When $\Delta=0$ (SSH-XX Hamiltonian), this well-known model, extensively studied in the literature, exhibits topological features such as localized edge states with eigenvalues within the gap for $\eta>0$. Moreover, some of its topological properties remain present even when a perpendicular anisotropy is introduced \cite{Li2018TopoPhase}. At the end of this work, we will also investigate the effects of a long-range dipolar interaction. The dipolar interaction is present in many physical systems, such as atoms and molecules on surfaces. We model the dipolar interaction by introducing the following term to the original Hamiltonian:

\begin{equation}\label{eqhdip}
 \hat{H}_{dip}=\frac{1}{4}\sum_{i<j} \frac{K}{r_{ij}^3}\left (\vec{\sigma}_{i}.\vec{\sigma}_{j}-3(\vec{\sigma}_i.\hat{n})(\vec{\sigma}_{j}.\hat{n})\right),
\end{equation}

\noindent where $\hat{n}=\hat{y}$ as shown in Fig \ref{fig1}, $r_{i,i+1}=a$ is the distance between consecutive spins,  $K$ is de dipolar coupling, and $a$ is the lattice parameter. As we will see later, the inclusion of this last term has profound consequences on the dynamics of the system. We can see that the dipolar Hamiltonian does not commute with the total magnetization of the system ($[\hat{H}_{dip},\sigma_z]\neq 0$). That the total magnetization is not a conserved quantity means that, given an initial condition in a fixed magnetization subspace, the time evolution does not remain within this subspace, implying that for any calculation with dipolar interaction, we must consider the entire Hilbert space and not just the space where the initial condition lives.
We are also interested in the effect of a homogeneous external magnetic field, so we add to the total Hamiltonian the following term:

\begin{equation}
 \hat{H}_Z=\frac{1}{2}g\mu_B B_z\sum_j^N \sigma_j^z,
\end{equation}

\noindent where $\mu_B$ is the Bohr magneton and $g$ is the giromagnetic factor.

Through the paper, we use units such that $\hbar=1$, $J=1$, and $K/a^3 = 0.1 \times J$. These choices imply that energy and time are dimensionless quantities.

\subsection{Autonomous Transmision}
\label{sat}
\subsubsection{One Excitation}

The Hamiltonian in Eq.~(\ref{eqhssh}) commute with the $z$-component of the total magnetization of the system, so it is natural to study the dynamics of quantum states restricted to subspaces with a fixed number of excitations. When we add the dipolar Hamiltonian Eq.~(\ref{eqhdip}), we must be careful and take into account the complete Hilbert space to study the dynamics. It is simple to show that the fidelity of the transmission of arbitrary one-qubit quantum states depends only on the transition probability
\begin{equation}
 \left|f_{AB}(\tau)\right|^2 = |\left\langle \mathbf{1}_B \right| U(\tau) \left|  \mathbf{1}_A \right\rangle |^2,
\end{equation}
where the one-excitation states $\left|  \mathbf{1}_A \right\rangle$, $\left|  \mathbf{1}_B \right\rangle$ correspond to a single excitation localized in the $A$ or $B$ sites of the chain, respectively.  The time-evolution operator, $U(\tau)$, is given by
\begin{equation}
U(\tau) = \exp{(-i H \tau)} ,
\end{equation}

\noindent where $H$ is the Hamiltonian, and $\tau$ is the time elapsed since the excitation dwelled in site $A$. In this work, we will be interested in transmitting one excitation between the ends of the chain, that is, we are interested in $P_1(\tau)=\left|f_{1N}(\tau)\right|^2$. Often, the performance of the protocol used to transfer arbitrary one-qubit quantum states is analysed using the averaged transmission fidelity.

\begin{equation}
F_1(\tau) = \frac{1}{2} + \frac{\left|f_{1N}(\tau)\right|}{3} \cos{\gamma}+\frac{\left|f_{1N}(\tau)\right|^2}{6} ,
\end{equation}

\noindent where the value of the phase $\gamma$ depends on particular values of the eigenvalue spectrum. Nevertheless, applying an appropriate magnetic field, the phase factor becomes equal to unity.  Later on, we will present only results for the transmission probability, since the averaged fidelity is a monotonically increasing function of it.

\subsubsection{Two Excitations}

The case of two excitations is technically more complicated. We start by looking at something similar to what we observed in the case of a single excitation. We prepare two excitations at the first two sites at one end of the chain, let these two excitations evolve, and after a specific time, we expect them to arrive at the last two sites in the chain. We define the probability of transmission for two excitations as

\begin{equation}
 P_2(\tau) = |\bra{0,0,0,...1,,1}e^{-iH\tau/\hbar}\ket{1,1,0,0,0,...}|^2= |f_{12,N-1N}(\tau)|^2. \label{eqf2e1}
\end{equation}
While the two-excitation transmission probability has information about the time evolution of the system, the practical problem involves the transmission of general two-qubit states, which are given by

\begin{equation}
\ket{\phi}=a\ket{00}+b\ket{10}+c\ket{01}+d\ket{11} ,
\end{equation}

\noindent where $a,b, c$, and $d$ are complex coefficients such that $||\ket{\phi}||=1$, and the kets correspond to chain states without excitations, with a single excitation located on the first or second qubit, and with two excitations. As shown in Reference \cite{Apollaro2022}, it is straightforward to show that the averaged fidelity for the transmission of arbitrary two-qubit states, $F_2$, is given by

\begin{eqnarray}
 &&F_2(\tau) = 0.25 + (|f_{1,N}|^2+|f_{2,N-1}|^2+|f_{12,N-1N}|^2)/20+ \nonumber \\ &&Re(f_{1N}+f_{2,N-1}+f_{12,N-1N}+(f_{2,N-1}+f_{12,N-1N})f_{1N}^*+f_{12,N-1N}f_{2,N-1}^*)/10 . \label{eq2e3}  
\end{eqnarray}

Some authors, see Reference~\cite{Wang2022}, also consider the transmission of entangled two-qubit states that are linear superpositions of one-excitation states, as, for instance, the state $\ket{\phi_{12}}=b\ket{01}+c\ket{10}$. The averaged transmission fidelity for these states is given by

\begin{equation}
 F_{12}(\tau)=\frac{1}{3}(|f_{1,N-1}|^2+|f_{2,N}|^2+|f_{2,N-1}|^2/2+|f_{1,N}|^2/2)+\frac{1}{3}Re(f_{1,N-1}f_{2,N}^*). \label{eq2e2}
\end{equation}

\subsection{Optimal control of the state transmission}
\label{soct}

In general, Heisenberg chains do not present perfect transmission, or even worse, the fidelity of transmission is so low at any time that it is of no practical interest. In some cases, optimizing the coupling coefficients removes this problem \cite{Serra2022PLA,Ferron2021UnderstandingTP,Serra2022JPA}. If constructing a device with its coupling coefficients tuned to precise values is not feasible, using external control means is advisable. We obtain appropriate control pulses that maximize the probability, or the fidelity of transmission, using Optimal Control Theory (OCT) [\onlinecite{Werschnik2007}]. OCT has a history of success dealing with different situations in the quantum information arena. Nevertheless, it is worth mentioning the 
growing literature dealing with controlling the time evolution of quantum protocols by applying control pulses designed using machine learning methods. Reinforced learning is one of the most employed and has been used to assess the quantum speed limit \cite{Zhang2018}, \cite{Bukov2018}, quantum state preparation \cite{Sivak2022, Porotti2022}, and the design of control gates \cite{Niu2019}. 

Optimal Control Theory requires selecting the number and type of actuators to drive the time evolution of the system, together with a cost functional to be optimized. In this study, we consider only one actuator, which is a time-dependent magnetic field applied to the first site of the spin chain. The cost functional generates a variational problem, solved by the iterative algorithm of Krotov. Krotov’s algorithm is a gradient-based optimal control method in which the dynamics of the quantum system are steered by minimizing a cost functional. For state-to-state transfer, the final-time part of the functional is the infidelity,
\begin{equation}\label{J1}
J_T[\Psi(T)]=1-|\langle \Psi_{target}|\Psi(T)\rangle|²,
\end{equation}

\noindent where $|\Psi_{target}\rangle=|\mathbf{1}_N \rangle$  in the case of a single excitation at site $N$. The complete cost functional additionally includes a term ensuring consistency with the Schrödinger dynamics, and a running cost that penalizes undesirable control fields, such as abrupt switching. The Krotov update equations follow from the variational minimization of this functional, guaranteeing monotonic convergence of the infidelity \cite{coden2024quantum}.

\subsection{Systems with Static Disorder}
\label{smdis}
We know that the perfect engineering of all spin couplings is complex to achieve in an actual experimental implementation, and it is interesting to analyze the performance of different spin chains related to the imperfect construction
of such devices from a theoretical point of view.
The robustness of quantum state transfer in spin chains under static disorder has been extensively studied in the literature \cite{Serra2022JPA, Zwick2015,Stolze2014,Petrosyan2010}. In this context, static disorder is a consequence of the construction procedure of the system that we modeled as a spin chain. We can model the static perturbations by adding a random term proportional to the coupling strength.

\begin{eqnarray}\label{disorder}
J_i \rightarrow J_i (1 + D_J \xi_{J_i}),\\ \nonumber
K_i \rightarrow K_i (1 + D_{K}\xi_{K_i}),
\end{eqnarray}

\noindent where each $\xi_i$ is an independent uniformly distributed random variable in the interval $[-D, D]$ and $D$ is a positive real number that characterizes the maximum perturbation strength relative to $ J_i$ or $K_i$. The disorder strongly depends on the experimental method used to build the spin chains. When we add static disorder, the transmission probability or the fidelity averaged over realizations of the disorder is introduced as a figure of merit:

\begin{eqnarray}
 \bar{P} = \left\langle P \right\rangle_{\xi} \\ \nonumber
 \bar{F} = \left\langle F \right\rangle_{\xi}
\end{eqnarray}

\noindent where the average is a simple equally weighted average. This scenario means that, for a chain with $N=8$ spins, a single realization of the disorder involves sorting $N-1$ random numbers $\lbrace\xi_i^j\rbrace$, for $1\leq i \leq N-1$, one for each exchange coupling, and the supraindex $j$ is such that $1\leq j \leq M_r$. $M_r$ is the number of realizations. For fixed values of $\eta$, $\Delta$ and $K_i$ the transmission
probability for an arrival time $T$, we calculate

\begin{equation}
 \bar{P}_D(T) = \frac{1}{M_r} \sum_{j=1}^{M_r} P_{\xi^j(D)}(T),
\end{equation}

\noindent where $P^j_{\xi(D)}(T)$ is the probability of transmission obtained for the $j$-th realization of the disorder. Later, we will present results for $\bar{P}_D(T)$, where $T$ is the arrival time for systems without disorder.


\section{Autonomous Transmission of One Excitation in a Spin Chain}
\label{srat1e}
In this section, we analyze the autonomous evolution of the system, observing the free evolution of a single excitation and identifying the maximum probability of transmission within a specified time interval \cite{bose2003quantum}. We aim to understand how the chain transmits information for different values of the anisotropy and the dimerization parameters. Of course, we are also interested in how robust the transmission is when there are defects in the engineered design of the spin chain.


\subsection{Efficient transmission of one excitation}
\label{srat1e-ef}

The autonomous transmission of one excitation through a chain with the Hamiltonian in  Eq. (\ref{eqhssh}) is the subject of this subsection. Changing the parameters $\eta$ and $\Delta$ modifies the spectrum and eigenvectors of the Hamiltonian. Our purpose is to identify sectors of the parameter space where the autonomous time evolution leads to high-quality quantum state transfer. Let us remember that the region where $\eta>0$  has more or less pronounced topological properties, such as localized states and a non-trivial Berry phase \cite{Li2018TopoPhase}. The region where $\eta<0$ does not show anything resembling a topological trait. From now on, we will denote the former region as the topological-like region, while the latter will be called the non-topological region. 

Fig. \ref{fm1} shows the contour map of the maximum probability reached in a time interval $[0,2000]$ as a function of the dimerization parameter and the perpendicular anisotropy for four chains of different lengths. The first thing we observe is relatively trivial. As the chain gets longer, transmitting an excitation becomes more complicated. However, we can see that for some values of $\Delta$ and $\eta$, the transmission of one excitation remains efficient, even for long spin chains. The second thing we note is related to the convoluted patterns shown. There is a bat-shaped region where the transmission probability can reach very high values in the time window explored. Outside this region, the characteristic times required to achieve high enough values of the transmission probability are considerably longer, as shown in Fig. \ref{figevn8}, causing the first maximum of the transmission probability to fall outside the time window. As we increase the time interval, the bat-shaped region expands and shows a bigger density of good values. 
In the region where $P$ reaches high values, the transmission patterns appear highly complex and complicated to interpret. We observe that tiny variations in $\Delta$ and $\eta$ can lead to abrupt transitions between efficient and poor transmission, except in certain areas, mainly located in the non-topological regime or on the borders of the bat-like region at the top of the maps in the topological-like region.

A key feature of the transmission maps is their distinct structure in the region where $\eta>0$. In Fig. \ref{edn10}, we show the degree of edge localization $\chi$ (left panel), which is given by:

\begin{figure}[h]
    \centering
    \includegraphics[width=0.75\textwidth]{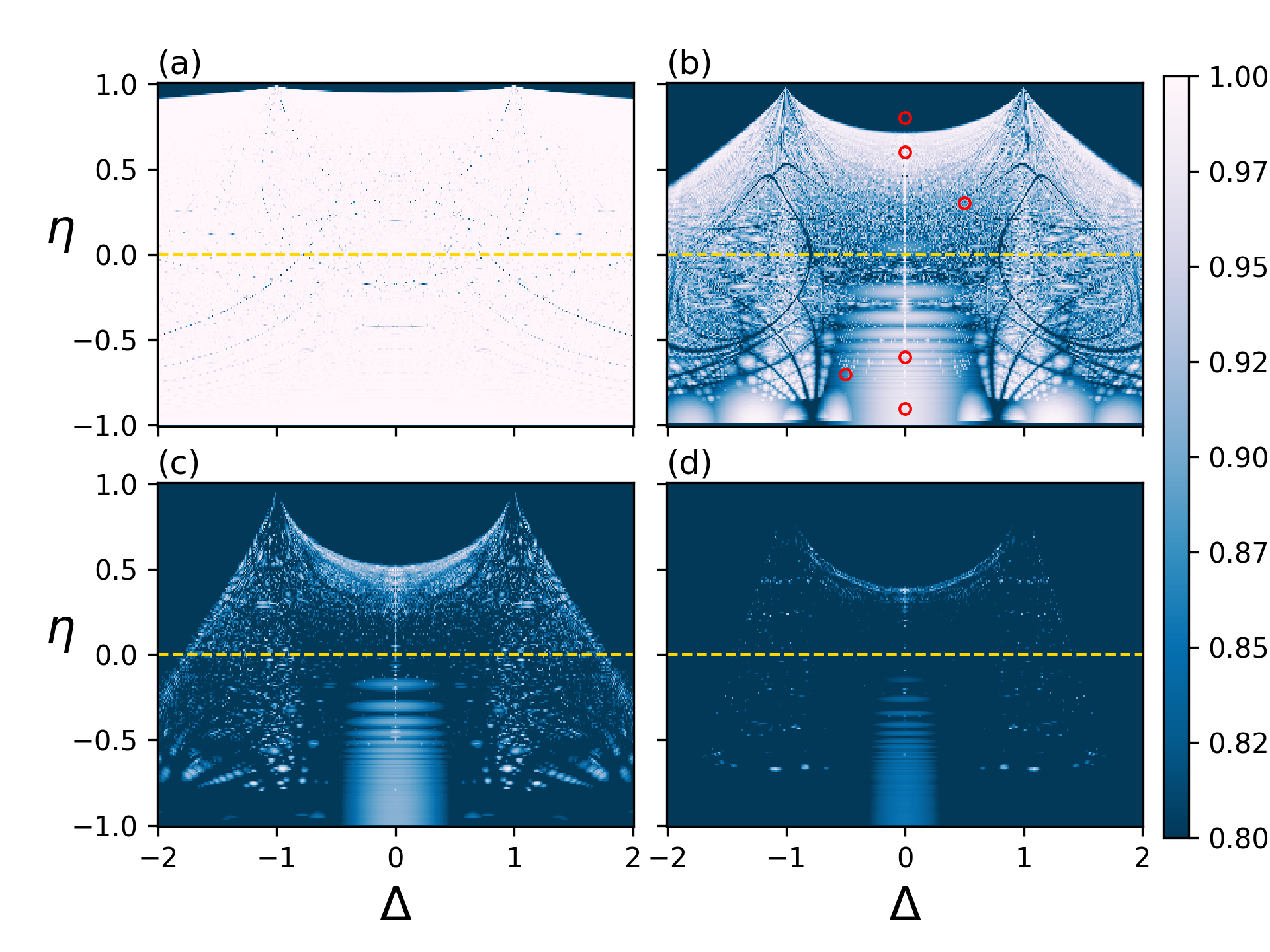}
    \caption{Contour plots for the maximum transmission probability value in a time interval $[0,2000]$ as a function of ($\Delta$, $\eta$) for chains with different lengths. (a) $N=4$, (b) $N=8$, (c) $N=12$, and (d) $N=16$.}
    \label{fm1}
\end{figure}

\begin{equation}\label{eqchi}
    \chi=\sum_{i=1}^2\left|\langle\Psi_{k_i}\left|\right.1\rangle\right|^2\;, 
\end{equation}

\noindent where $k_i$ corresponds to the two largest values of $|\langle\Psi_{k_i}\left|\right.1\rangle|^2$. To obtain the two largest values, it is necessary to calculate the projection over the whole set of eigenvectors. A value of $\chi=1$ would indicate that the entire projection lies in two edge states with probability $1/2$ each. Fig. \ref{figlocN10} (a) shows the map of  $\chi$ as a function of $\eta$ and $\Delta$. The region where $\chi>0.8$ shows that the system exhibits highly localized edge states. Note that around $\Delta= \pm 1$, where the chain is SSH-Heisenberg-like, there are no edge states for any value of $\eta$. The high localization regions coincide, as seen in Fig. \ref{figlocN10} (b), with the areas where it is challenging to reach high probability values in reasonable times. 

In regions with highly localized edge states, the coefficients corresponding to these states determine the dynamical behaviour. As a result, the probability exhibits a simple harmonic form,

\begin{equation}
P(t) \approx \sin^2\left( \frac{\varepsilon t}{2} \right),
\end{equation}

\noindent with $\varepsilon=E_{k_2}-E_{k_1}$. In the topological regime, this energy splitting decreases exponentially with $\eta$, explaining the exceedingly slow oscillations outside the bat-like region. Some authors point to this region as a scenario where exploiting the protection associated with topological properties is advantageous. Avoiding decoherence effects over such prolonged intervals is the object of many studies.

In Fig. \ref{figevn8}, we plot the evolution of the transmission probability $P_1(t)$ for six different sets of parameters for a chain length $N=8$, identified with red circles in Figure \ref{fm1} (b). The curves in Fig.~\ref{figevn8} show different dynamic behaviors that arise as we explore the space of parameters $\Delta$ and $\eta$. In the topological-like region (Fig. \ref{figevn8} (a)), the probability evolution exhibits slow dynamics for $\eta$ large, compatible with the discussion of Fig. \ref{figlocN10} (a). The inset zoom shows a closer view of the oscillatory pattern, with some interference features and also making the increasing period of oscillations more evident as $\eta$ changes from $0.5$ (orange line) to $0.8$ (green line). In contrast, the time evolution of the transmission probability obtained for parameters chosen from the non-topological region (Fig. \ref{figevn8} (b)) displays a different behavior. The probability shows distinct patterns, with some cases appearing to reach high values or peaks rather quickly. However, considering that most of these chains present pretty good transmission (PGT) \cite{serra2023scaling,serra2025perfect}, it is reasonable to expect that, given sufficient time, many of these cases will eventually reach even higher transmissions. The detailed oscillatory features shown in the inset reveal that the parameters chosen produce similar dynamical behavior.

To gain a better understanding of the patterns observed on the transmission probability maps, we will analyze the case of the chain with $N=4$ in the subspace of one excitation, which can be solved analytically. The exact eigenvalues of the $4\times4$ matrix are

\begin{equation}
\lambda_1 = -\frac{1}{4}(D_{-}+\nu_+), \quad  
\lambda_2 = -\frac{1}{4}(D_{+}-\nu_+), \quad  
\lambda_3 = \frac{1}{4}(D_{-}-\nu_+), \quad  
\lambda_4 = \frac{1}{4}(D_{+} +\nu_+),
\end{equation}
 
\noindent where $\nu_\pm= 1\pm\eta$, and $D_{\pm} = \sqrt{4\nu_-^2 + (\Delta \pm 1)^2\,\nu_+^2}$. Unlike the XX case, no simple symmetry relations hold among the eigenvalues. The transmission probability is given by

\begin{equation}\label{ec-P1N4}
P_1(t) = 4\nu_-^4 \left| \sum_{i=1}^{4} (-1)^i \frac{e^{-i\lambda_i t}}{4\nu_-^2+(4\lambda_i+\nu_+\,\Delta)^2} \right|^2.
\end{equation}

\begin{figure}[h]
    \centering
    \includegraphics[width=0.75\textwidth]{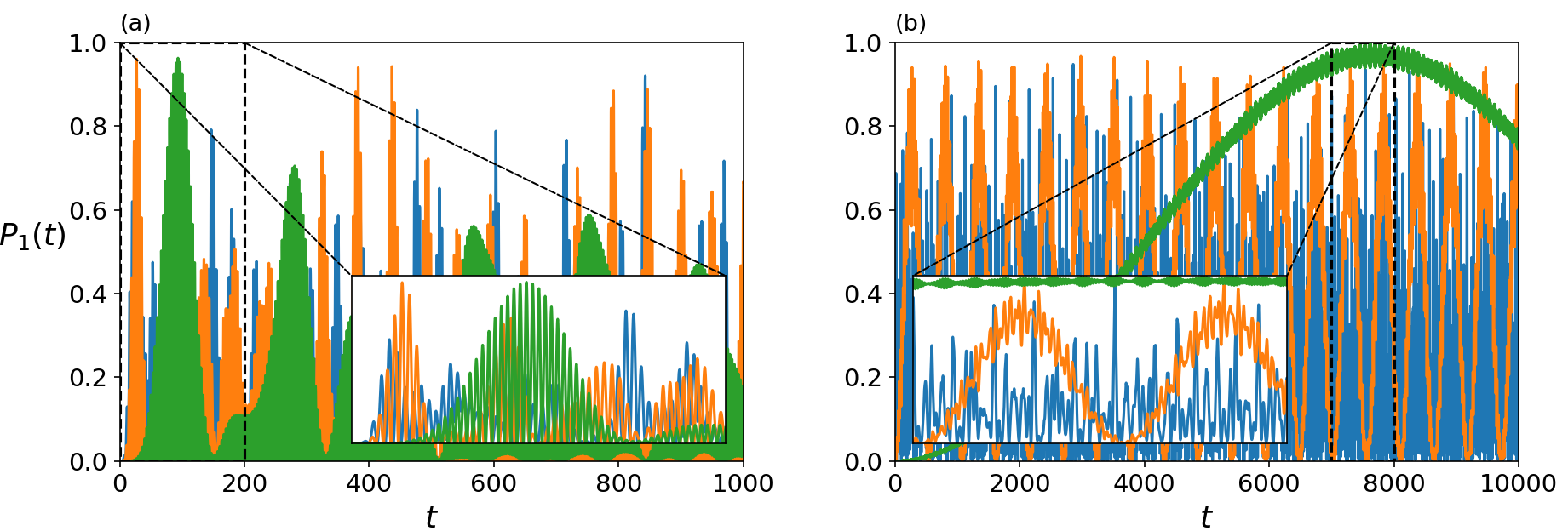}
      \caption{Transmission probability $P_1(t)$ as a function of time for $6$ different parameter sets for a spin chain (Eq. (\ref{eqhssh})) of length $N=8$. The $\eta$ and $\Delta$ values corresponding to these curves are shown as red circles in Fig. \ref{fm1} (b). Panel (a) shows the transmission probability for $\Delta=1$ and $\eta=-0.5$ (Blue), $\Delta=0$ and $\eta=-0.6$ (Orange), and $\Delta=0.15$ and $\eta=-0.88$ (Green). We plot in panel (b) the transmission probability for $\Delta=0.9$ and $\eta=0.4$ (Blue), $\Delta=0.5$ and $\eta=0.6$ (Orange), and $\Delta=0$ and $\eta=0.78$ (Green).} \label{figevn8}
\end{figure}

  \begin{figure}[h]
    \centering
    \includegraphics[width=0.75\textwidth]{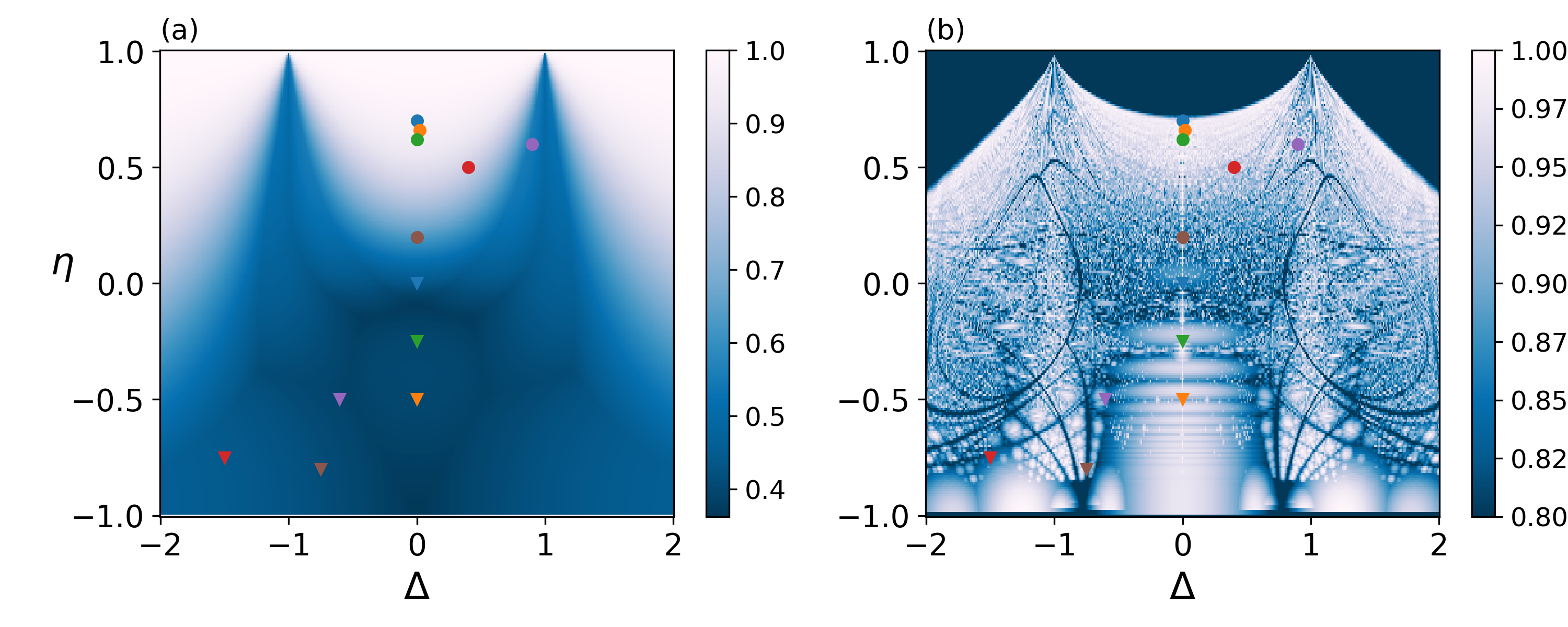}
    \caption{(a) Degree of localization of the wave function at the edges of the chain (b) Contour Maps ($\Delta$, $\eta$) for the maximum probability $P_1$ in a time interval $[0,2000]$. Data in both panels corresponds to a spin chain of length $N=8$ \label{edn10}    }
    \label{figlocN10}
\end{figure}

\begin{figure}[h]
    \centering
    \includegraphics[width=0.75\textwidth]{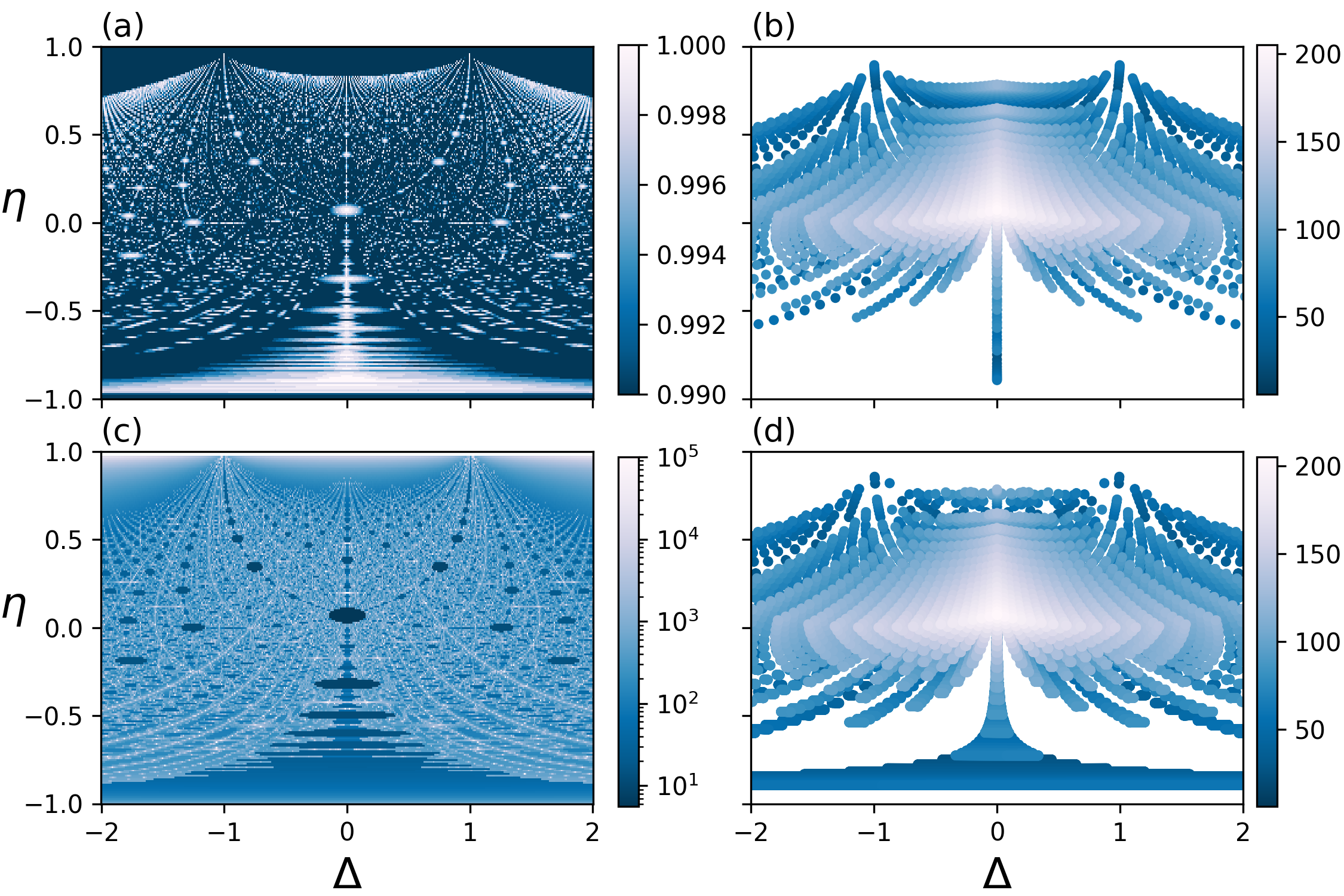}
    \caption{Exact calculations for $N=4$. (a) Maximum value of $P_1$ in a time interval $[0,200]$ as a function of $\eta$ and $\Delta$. (b) Values of $\eta$, $\Delta$ and $T$ for which the Kay condition \cite{kay2022PTheisen} is fulfilled for $q\leq 37$. (c) Time needed by $P_1$ to reach values greater than $0.99$.
(d) Values of $\eta$, $\Delta$, and $T$ that allow the eigenvalues to fulfill Kay's condition with a tolerance $\epsilon=0.001$ and for $q\leq 37$. }
    \label{fign4}
\end{figure}

\noindent The expression in Eq.~\eqref{ec-P1N4} does not have a simple form, but a numerical evaluation, plotted in Fig. \ref{fm1} (a), shows regions in $(\Delta, \eta)$ where $P_1(t)$ reaches high values of the transmission. In Fig.~\ref{fign4}, we present a series of results that help clarify some patterns observed in Fig.~\ref{fm1}. Figure \ref{fign4} (a) depicts the same data as Fig. \ref{fm1} (a) but over a smaller time interval, with a maximum time of $200$. This figure permits a visualization of the regions where transmission is particularly efficient. Notably, we observe disks where the transmission probability for short times is exceptionally high and insensitive to minor variations of the Hamiltonian parameters. Having the expression for the eigenvalues, we can ask them to satisfy the condition of ordered \cite{kay2022PTheisen} or partially ordered spectrum \cite{Ferron2021UnderstandingTP} to find the values of the parameters for which there exists perfect or almost perfect transmission. Following \cite{Ferron2021UnderstandingTP} and \cite{kay2022PTheisen}, we expect near-perfect QST in Heisenberg spin chains whose spectrum satisfies. 

\begin{equation}\label{kai}
\delta_i=\lambda_{i+1} - \lambda_i \sim q_i\pi/T
\end{equation}

\noindent where $q_i$ is an odd natural number. Perfect transmission appears when Eq.~\ref{kai} is an equality. For the case $N=4$, this condition gives us three equations to determine the values of $\eta$, $\Delta$, and the arrival time $T$. Fig. \ref{fign4} (c) shows the time required to reach a target transmission probability $P>0.99$ as a function of the Hamiltonian parameters. The structure of this map is similar to that of Fig. \ref{fign4} (a). Importantly, we observe that in some regions, the system reaches high values of $P$ within relatively short times, indicating favorable conditions for fast state transfer. The most insightful information comes from Fig. \ref{fign4} (b), where we plot the solutions to the set of equations derived from the Kay  condition \cite{kay2022PTheisen}, Eq. (\ref{kai}), in the case that equality is fulfilled and evaluated for odd integer numbers $q \leq 37$. These points explain several of the features observed in Fig. \ref{fign4} (a). In particular, the lines without points in the topological-like regime align with the dark fringes observed in the probability map, where transmission remains low. Conversely, points that satisfy the Kay condition correlate with some of the bright spots in Fig. \ref{fign4} (a), confirming the perfect transmission (PT) for these points. In the non-topological region, points satisfying the Kai condition are less frequent, contradicting the bright horizontal bands in the probability contour map. We note that the $XX$ chain exhibits PT for many values of the dimerization parameter $\eta$ with the times considered here. To analyze regions with high or almost perfect transmission, we relaxed the Kay condition and plotted the values of $\eta$ and $\Delta$ that satisfy $q_i\pi/T-\epsilon \leq  \delta_i \leq q_i\pi/T+\epsilon$ with a tolerance $\epsilon=0.001$ in Fig. \ref{fign4} (d). This map identifies areas of high transmission in the non-topological regime, mainly for the horizontal bright bands for $\eta\leq -0.5$.

Regrettably, the procedure used for the detailed analysis carried out for a chain with four spins is simply not doable for longer chains, since there are no analytical solutions for the roots of characteristic polynomials of higher degree. Nevertheless, it is clear that the bright spots in Figures~\ref{fm1} and \ref{figlocN10} (b) correspond to parameter regions where Kay's conditions are exactly or almost exactly satisfied. The size of the spot around which a point where an exact solution happens strongly depends on the chain length and the extension of the temporal window used to explore the time evolution of $P_1$. Indeed, spotting a point on the $(\eta,\Delta)$ plane where the transmission probability is good enough is an increasingly difficult task for increasing chain lengths, except near the axis where $\Delta=0$ and the arch that delimits the bat-like region above. Moreover, the tuning of the $\eta$ and $\Delta$ values needed to achieve higher values for the transmission probabilities is more demanding on the $\eta>0$ region than on the non-topological region.


\subsection{Disorder effect}
\label{srat1e-de}

Studying how static disorder in the coupling coefficients affects the transmission of one excitation is the subject of this Section. We study the transmission efficiency in chains whose coupling coefficients differ from their ideal values due to the presence of static disorder. We consider that the disorder strength, $D$,  goes from $1\%$ to $10\%$. We will present results obtained by averaging over $100$ independent disorder realizations, which is enough to get statistically sound results. We first compute the maximum transmission probability $P_1$ and the time we reach this probability for each pair of $\eta$ and $\Delta$ in the absence of disorder. Then, using those same parameters and the calculated times, we evaluate the evolution for different amplitudes of disorder.

\begin{figure}[h]
    \centering
    \includegraphics[width=0.75\linewidth]{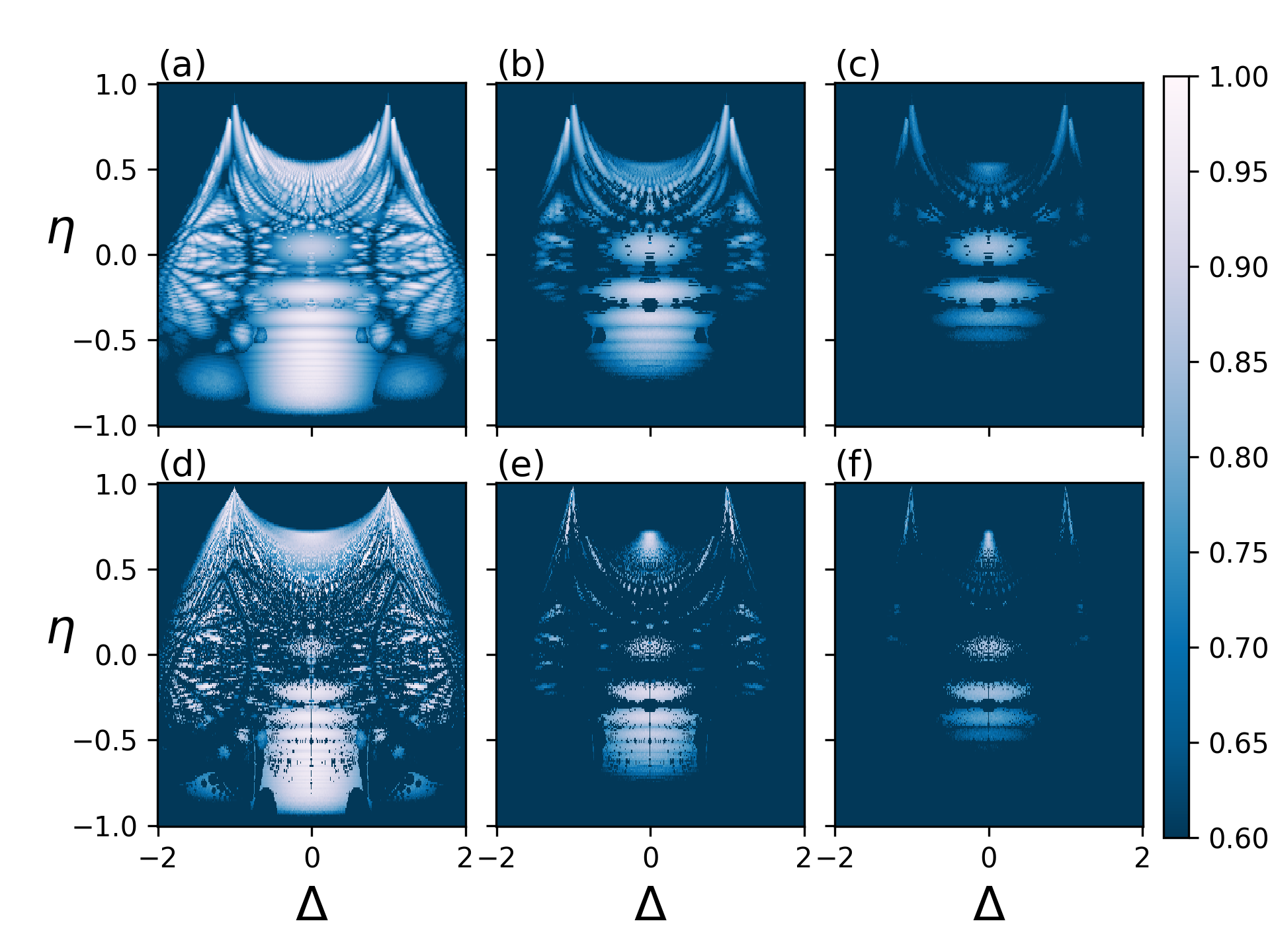}
    \caption{Contour map for the maximum transmission probability of an 8-site spin chain with disorder. The top panels show the results for the optimal time found within the interval $[0, 200]$, while the bottom panels correspond to the optimal time found within $[0, 2000]$. From left to right, the panels correspond to different disorder amplitudes: (a) and (d) $D = 0.01$, (b) and (e) $D = 0.05$, and (c) and (f) $D = 0.1$}
    \label{disorder1}
\end{figure}

Fig. \ref{disorder1} presents six panels showing the impact of increasing disorder amplitude on the maximum transmission probability $P_1$ as a function of the Hamiltonian parameters $\Delta$ and $\eta$ for $N=8$. Panels (a), (b), and (c) correspond to a maximum evolution time of $T=200$, while panels (d), (e), and (f) extend the time window to $T=2000$. As the disorder is present, we allow that the maps on each panel show areas where the average transmission exceeds $P_1>0.6$. Otherwise, since noise impairs the transmission, choosing a larger reference value would produce an almost completely dark map. 

As expected, Fig.~\ref{disorder1} shows that the transmission quality worsens with increasing disorder. A closer comparison between the upper panels and the lower ones reveals a nontrivial behavior: in the non-topological region, as shown in the previous Section, transmission reaches good values for relatively short times, and allowing the system to evolve longer exposes it to the destructive effects of disorder. In contrast, in the $\eta>0$ region, characterized by slower dynamics mainly driven by edge states, the transmission probability is more resilient to disorder.
This contrast shows a difference between the two transmission mechanisms. If the purpose is to transmit quickly, we will prefer to be in the non-topological region. If we had to wait longer transmission times, working with chains where the edge states control the dynamics would be a better choice.

\begin{figure}[h]
    \centering
    \includegraphics[width=0.75\linewidth]{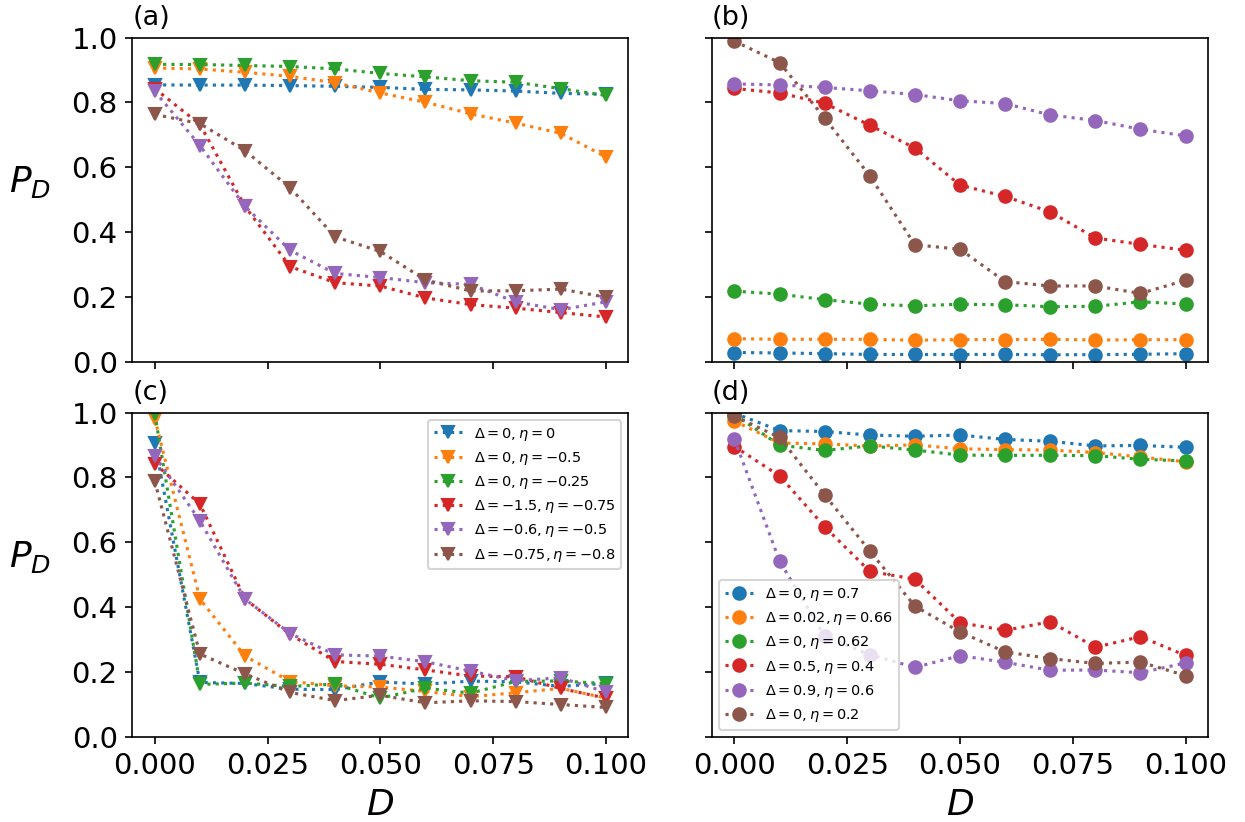}
    \caption{The four panels show the averaged transmission probability as a function of the disorder strength. We choose the Hamiltonian parameters from the two possible regions as shown by the color points in Fig. \ref{edn10} (we use the same color code and point shape in these curves in order to identify parameters $(\eta,\Delta)$), and the time window for each panel as follows: (a) Non-topological region and $T=200$, (b) topological-like region and  $T=200$, (c) Non-topological region and $T=2000$, (d) topological-like region and  $T=2000$.}
    \label{disorder2}
\end{figure}

Figure \ref{disorder2} provides a detailed view of the effect of disorder for an $N=8$ spin chain on some specific points of the parameter space represented by colored dots in Fig. \ref{figlocN10}. The figure has four panels: the top panels ((a) and (b)) correspond to short evolution times, while the bottom panels ((c) and (d)) represent long evolution times. Panels on the left ((a) and (c)) correspond to parameters chosen from the non-topological regime, and those on the right ((b) and (d)) to the topological-like one.

Figure \ref{disorder2} (a) shows the effect of disorder on the transmission probability in the non-topological region for short times for three representative points selected from the bright areas in the probability map. Notably, the green, orange, and blue curves maintain high transmission for disorder levels up to $5\%$, demonstrating a certain degree of robustness. In contrast, the remaining curves, corresponding to points outside the bright areas, exhibit an abrupt decay: even minimal disorder is enough to suppress transmission significantly. At longer times, as shown in Fig. \ref{disorder2} (c), the situation is dire for nearly all the selected points. Even the cases that work relatively well for short times are strongly affected by the disorder, which is consistent with the previous discussion regarding Fig. \ref{figevn8}. Points selected from the non-topological region achieve high transmission maxima at very short times. As a result, when we examine the system transmission abilities at short times, it is clear that the disorder does not have enough time to spoil the transmission. In contrast, at longer times, the effect of disorder becomes considerably more destructive.

The behavior in the $\eta>0$ region is quite different. In Fig. \ref{disorder2} (b), for short times, the system shows poor transmission for large values of $\eta$, even without disorder. On the other hand, points farther from the strongly localized regime exhibit high transmission without disorder, but their performance rapidly degrades with the introduction of disorder. Figure \ref{disorder2} (d) shows a different picture of this behavior for long waiting times. The curves corresponding to points with large $\eta$ and $\Delta \simeq 0$, which exhibit poor transmission at short times, now show excellent transmission and remarkable robustness against disorder, demonstrating the resilience of edge-state-mediated dynamics. Conversely, in the points associated with nonlocalized states, chains exhibit good transmission abilities without disorder; however, disorder destroys their efficiency, even with small amplitudes.

The presence of disorder, attributable in practical implementations to defective manufacturing of the chain elements, seems to strengthen the case of using $XX$ chains or $XXZ$ chains with a small value of the anisotropy parameter $\Delta$, and with a negative dimerization parameter $\eta<0$ chosen well inside the non-topological region. Chains with a Hamiltonian with such parameters are highly efficient, show fast quantum state transfer, and are pretty robust against errors produced by defective manufacturing or poor design. Moreover, the tuning of the parameters for this type of chain allows a certain degree of errors without compromising the quality of the transmission.


\section{Autonomous Transmission of arbitrary two-qubit states in a Spin Chain}
\label{srat2e}

In this section, we analyze what happens when we want to transmit arbitrary two-qubit states through the kind of quantum channels considered in the previous Sections. As discussed in the Introduction, there is more than one quantity that we can compute in this context, and they do not necessarily contain the same information. We begin by focusing on three specific quantities, the transmission probability of two excitations prepared on one extreme of the chains, $P_2$, the averaged transmission fidelity for entangled states that are a linear superposition of one excitation states,  $F_{12}$, and $F_2$, the averaged transmission fidelity where the average runs over all the possible two-qubits states prepared on the first two sites of the chain (see Eqs. (\ref{eqf2e1}), (\ref{eq2e2}) and (\ref{eq2e3})). Note that the time evolution for each quantity occurs on different subspaces. For $P_2$, the time evolution occurs on the two-excitation subspace exclusively, for $F_{12}$ it occurs on the one-excitation subspace, and for $F_2$ it occurs on both subspaces. 

\begin{figure}[h]
    \centering
    \includegraphics[width=0.75\textwidth]{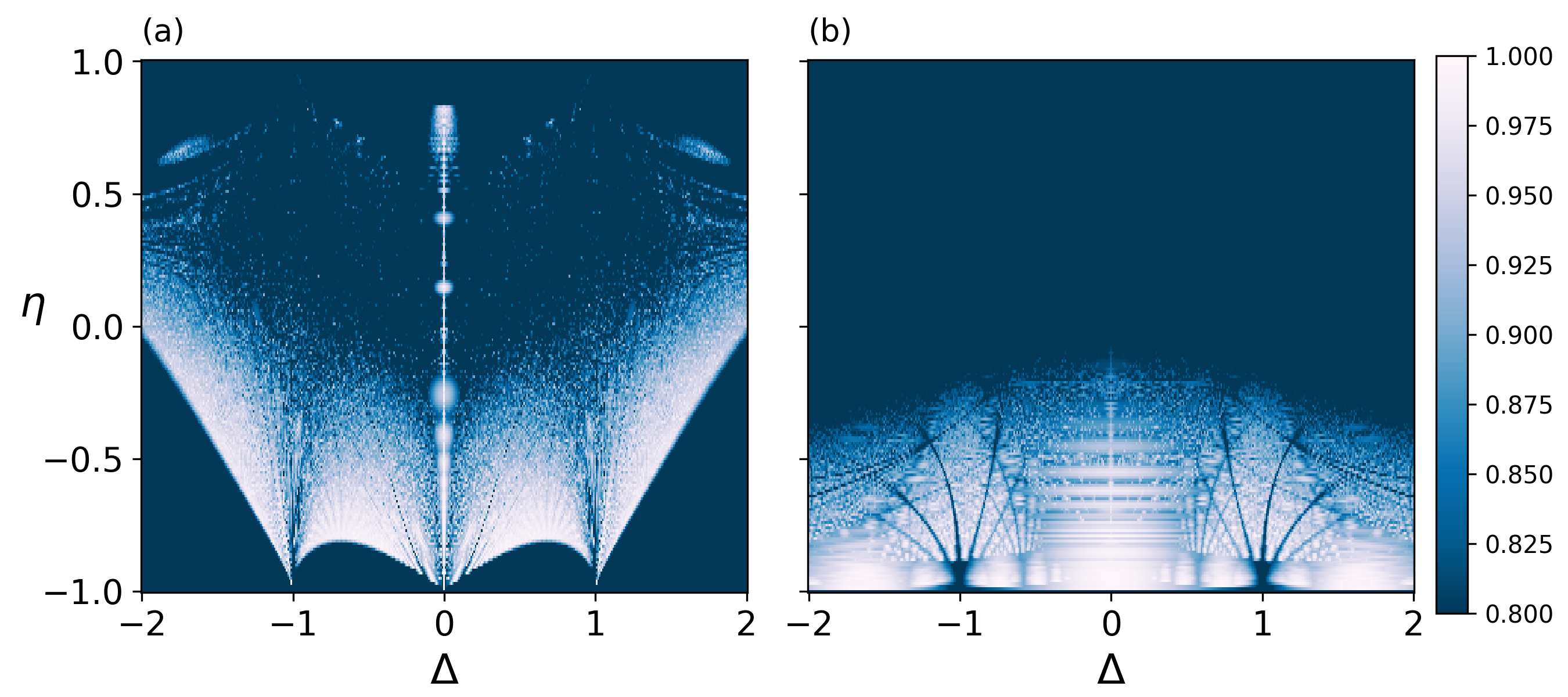}
    \caption{Countour plots ($\Delta$, $\eta$) for the maximum value of $P_2$ in (a) and $F_{12}$ in (b) for the Hamiltonian
    $\hat{H}_{ssh}$, an optimal time $t$ in $[0,2000]$ and a chain length of $N=8$.}
    \label{figmap2e-1}
\end{figure}

The quantity $P_2$ is the natural extension of what we have calculated so far in this work, and it measures the probability of successfully transferring two excitations from one end of the chain to the other. The second quantity, $F_{12}$, corresponds to the average fidelity of transmitting a two-qubit state of the form $\alpha|01\rangle + \beta|10\rangle$, averaged over all possible values $\alpha$ and $\beta$. Finally, $F_2$ gives the average fidelity for the transmission of arbitrary two-qubit states, accounting for the contribution of the one and two-excitation subspaces.

In Fig. \ref{figmap2e-1}, we present contour maps for the first two quantities discussed, $P_2$ and $F_{12}$, as functions of the dimerization and anisotropy parameters for a chain of length $N=8$. The resulting patterns differ significantly from one to another, and also from the characteristic bat-shaped structures observed in the single-excitation case.

For the simplest one, $P_2$, we observe a structure that retains some resemblance to the single-excitation maps. However, in this case, interesting features emerge everywhere, within both the topological and non-topological regimes. However, it is clear that for $\eta<0$ there are many more sets of parameters that produce high-quality transmission than for $\eta>0$. For positive and negative values of $\eta$, the $XX$ chain always shows good transfer abilities. 


\begin{figure}[h]
    \centering
    \includegraphics[width=0.45\textwidth]{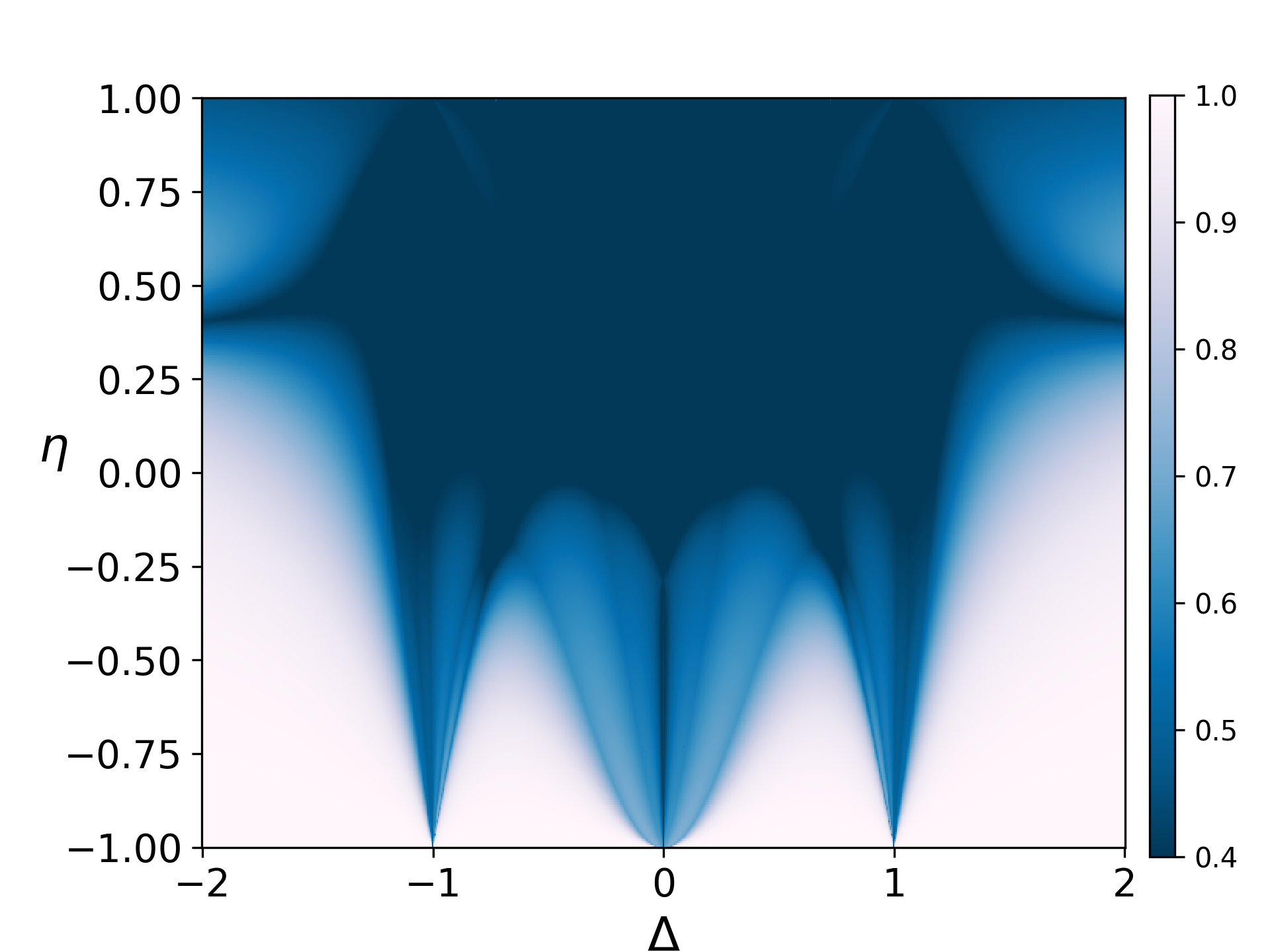}
    \caption{Countour plots ($\Delta$, $\eta$) for the degree of localization of two excitations at the edges of the chain for $N=8$.}
    \label{figloc2e}
\end{figure}

In Fig.\ref{figmap2e-1} (b), we plot the contour map for $F_{12}$ as a function of the Hamiltonian parameters. We observe that for $\eta > 0$, the maximum values of $F_{12}$ remain below 0.8, indicating that this region is not a good choice for transmitting quantum states of the form $\alpha|01\rangle + \beta|10\rangle$. One might believe that this is related to the waiting time, but this is not the case, and $F_{12}(t)$ never reaches high values when $\eta > 0$. On the other hand, for negative values of $\eta$, the structure of the map becomes similar to the one observed for the single-excitation case 
$P_1$. High values of $F_{12}(t)$ in this regime suggest that the excitation spends some time oscillating between the last two sites of the chain. We can understand this behavior by considering the structure of an SSH spin chain. For $\eta < 0$, the edges and their nearest neighbors interact strongly, and the chain becomes a sequence of weakly connected two-site cells. When the excitation reaches the last cell, it becomes temporarily trapped, oscillating there for a finite amount of time. 

As in the case of one excitation transmission, we also define the degree of edge localization for two excitations.

\begin{equation}\label{eqchi2}
\chi^{(2)} = \sum_{i=1}^{2} \left| \langle \Psi_{k_i} | 1,2 \rangle\right|^2 , 
\end{equation}

\noindent where $| 1,2 \rangle $ is the two-excitation basis state such that both excitations lie on the first and second qubit of the chain. Figure~\ref{figloc2e}, shows the contour plot of $\chi^{(2)}$ as a function of the chain parameters. Comparing with Figure~\ref{figmap2e-1} (a), it is appreciable that the whiter zones of Figure~\ref{figloc2e} close to $\eta=-1$, corresponding to higher values of the degree of localization, approximately match the darker zones of Figure~\ref{figmap2e-1} corresponding to zones with poor transmission of two-excitation states, measured with $P_2$ (see Eq.~\eqref{eqf2e1}). Notwithstanding this fact, part of the regions with higher values of $\chi^{(2)}$ are propitious for the transmission of entangled states, as shown in Figure~\ref{figmap2e-1} (b), as measured by $F_{12}$. At least for the window time considered, the transmission of one-excitation entangled states is appreciable only in the region $\eta<0$. Again, as is the case with the one-excitation transmission, good transfer abilities arise from the interplay between the fulfilment of Kay's conditions and localization at the edges. 

The analysis of $F_2$ is more complex. First of all, the symmetry around $\Delta = 0$, clearly visible in the earlier contour plots, breaks down in this case. To understand this behavior, let us recall the definition of $F_2$ and rewrite it in the following form:
\begin{equation}
        F_2 = \frac{1}{4} + (|f_1^N|+|f_2^{N-1}|+|f_{12}^{NN-1}|^2)/20 + R_f \label{eqrf}
\end{equation}

\noindent where 

\begin{eqnarray}
    R_f&=&|f_1^N|cos(\phi_1^N)+|f_{2}^{N-1}|cos(\phi_2^{N-1})+|f_{12}^{NN-1}|cos(\phi_{12}^{NN-1})+|f_1^Nf_2^{N-1}|cos(\phi_2^{N-1}-\phi_1^N) \\ \nonumber
    \label{eqrf2}
    &&+|f_1^Nf_{12}^{N-1N}|cos(\phi_{12}^{N-1N}-\phi_1^N)+|f_2^{N-1}f_{12}^{N-1N}|cos(\phi_{12}^{N-1N}-\phi_2^{N-1}). 
\end{eqnarray}

\noindent where we have expressed the coefficients $f_k^m$ in terms of their modulus and phase, as is standard for any complex number. As suggested by the results presented in previous sections, the moduli are completely symmetric with respect to the zero anisotropy axis. Then, there are the phases that introduce the asymmetry we have mentioned. This asymmetry can be controlled or eliminated by introducing a magnetic field in the perpendicular direction, $B_z$. 

\begin{figure}[h]
    \centering
    \includegraphics[width=0.75\textwidth]{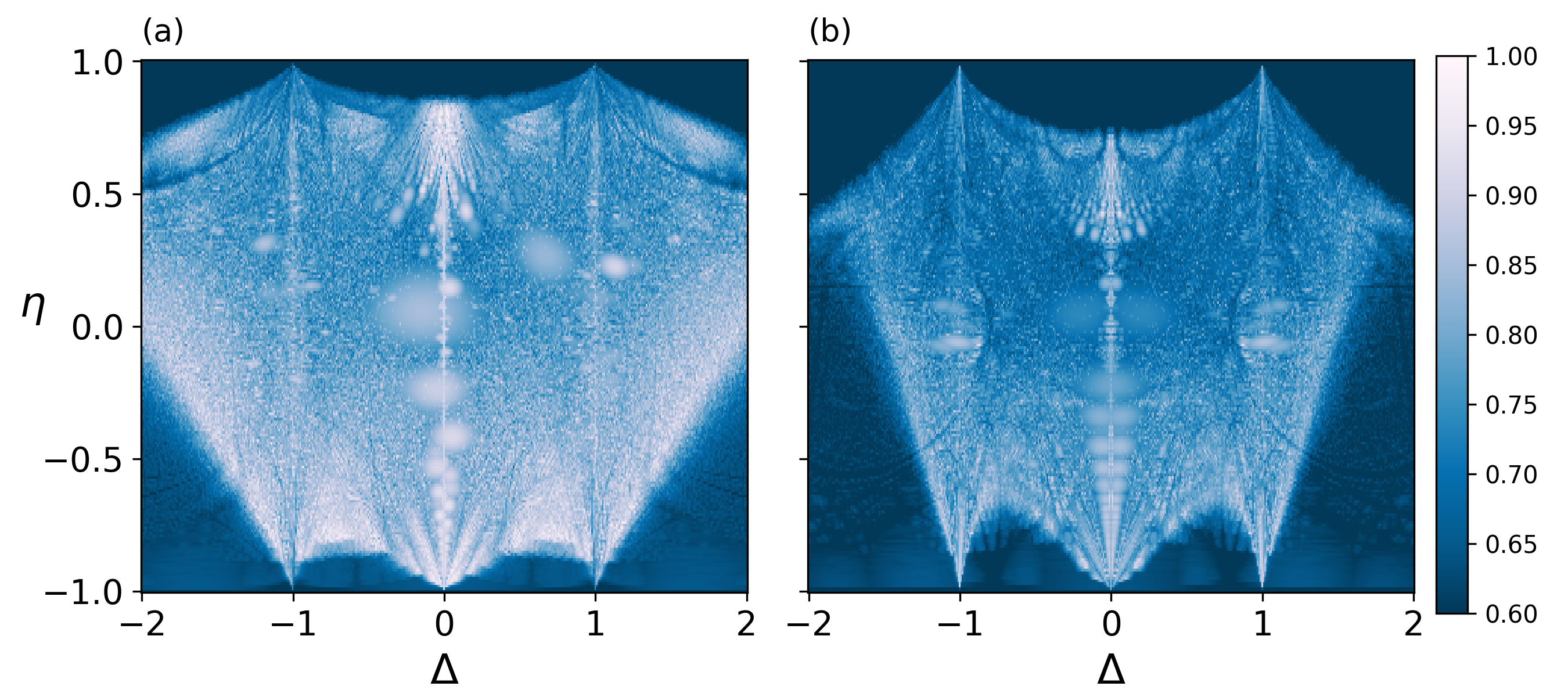}
    \caption{Countour plots ($\Delta$, $\eta$) for the maximum value of $F_{2}$ for the Hamiltonian
    $\hat{H}_{ssh}$, an optimal time $t$ in $[0,2000]$, optimal magnetic field and chains length of $N=6$ in (a) and $N=8$ in (b).}
    \label{figmap2e-2}
\end{figure}

The last term in Eq. (\ref{eqrf}) is the only one that can break the symmetry and is the only part of the expression that can be affected by the magnetic field.
By introducing this magnetic field, we obtain:

\begin{eqnarray}
    \tilde{R}_f&=&|f_1^N|cos(\phi_1^N+\gamma)+|f_{2}^{N-1}|cos(\phi_2^{N-1}+\gamma)+|f_{12}^{NN-1}|cos(\phi_{12}^{NN-1}+\frac{N-4}{N-2}\gamma)\\ \nonumber
    &&+|f_1^Nf_2^{N-1}|cos(\phi_2^{N-1}-\phi_1^N)
    +|f_1^Nf_{12}^{N-1N}|cos(\phi_{12}^{N-1N}-\phi_1^N-\frac{2}{N-2}\gamma)\\ \nonumber
    &&+|f_2^{N-1}f_{12}^{N-1N}|cos(\phi_{12}^{N-1N}-\phi_2^{N-1}-\frac{2}{N-2}\gamma)
    \label{Rtilde}
\end{eqnarray}

\noindent with $\gamma=(N-2)B_zt$. Now, the fidelity in the presence of a magnetic field, $\tilde{F}_2(t, B_z)$, depends on two free parameters, providing greater flexibility to optimize or constrain the fidelity according to specific requirements. For instance, if we are interested in it, it is possible to ask that the transmission protocol results symmetric with resepect to $\Delta=0$. This symmetrization would be achievable by finding the value of the magnetic field for which $\tilde{R}_f(+\Delta)=\tilde{R}_f(-\Delta)$. However, solving this equation analytically is quite complicated. We find that when $N/2$ is even, both $\tilde{R}_f(+\Delta)$ and $\tilde{R}_f(-\Delta)$ are the same periodic function of $\gamma$, but shifted by a phase of $(N/2 - 1)/\pi$. In contrast, the case when $N/2$ is odd is significantly more complicated, and the equality $\tilde{R}_f(+\Delta)=\tilde{R}_f(-\Delta)$ can only be achieved for specific values of $\gamma$ and should be solved numerically. 

Using the magnetic field as a control parameter to optimize the fidelity of state transmission seems to be a more interesting option. In the case of $N/2$ even, once we determine the optimal magnetic field for $\Delta>0$, we can calculate the corresponding field for $\Delta<0$ that not only optimizes the fidelity but also restores symmetry in the contour plot map. Unfortunately, for $N/2$ odd, this is no longer possible: there is no magnetic field that both optimizes the fidelity and symmetrizes the results around the axis $\Delta=0$. Fig. \ref{figmap2e-2} (a) shows the fidelity map for $N=6$, where we optimize both the waiting time and the magnetic field for each point in $(\eta,\Delta)$ space. In this case, we observe small asymmetries between the $\Delta > 0$ and $\Delta<0$ regions. In contrast, Fig. \ref{figmap2e-2} (b) shows the results for $N=8$, where we tune the magnetic field not only to maximize fidelity but also to force symmetry across the $\Delta=0$ axis.

Although optimizing over two variables, time and magnetic field strength, always provides larger values for the transmission probability than optimizing it only on time, the quality of the transmission does not increase significantly. For given pairs of the Hamiltonian parameters $(\Delta,\eta)$, the increase of the transmission probability value reaches an improvement of around $10\%-20\%$. Since the increments start from very low values of the transmission probability, the inclusion of a constant magnetic field does not change the overall assessment that this type of chain has poor transmission abilities when dealing with the transmission of arbitrary two-qubit states.

We do not include further figures concerning the results found for the behavior of $F_{12}$ and $P_2$ when static disorder is present. Nevertheless, a comment is in order. Adding static disorder always spoils the quality of the transmission. Still, when considering the transmission of arbitrary two-qubit states or one excitation entangled states, the decay observed is even more severe than the one observed in the case of the transmission of one-qubit states, irrespective of the region of the parameter plane under consideration.  


\section{Adding a dipolar term to the dimerized chain}
\label{srdip}

In this section, we include the dipolar interaction term (see Eq. (\ref{eqhdip})) in the chain Hamiltonian, which is common in several systems such as atomic chains. As discussed in [\onlinecite{coden2024quantum}], the presence of dipolar coupling worsens quantum state transmission, making the process significantly more difficult. The worsening comes from the fact that the dipolar interaction does not commute with the total magnetization of the system. As a result, there is no conservation of the number of excitations during the evolution of the system. In all the cases considered until here, the system evolved within a fixed excitation subspace determined by the initial state: one or two excitations prepared at one end of the chain remained confined to the corresponding subspace throughout the evolution. The wavefunction at a given time is now a superposition of states with weight in several subspaces of the Hilbert space, making it harder to recover one or two excitations at the opposite end of the chain, since the system explores subspaces that are orthogonal to the initial condition. However, as shown in [\onlinecite{coden2024quantum}], we can mitigate this problem by applying a magnetic field along the $z$-direction. This field penalizes transitions between different excitation subspaces, inducing an effective conservation of the excitation number. Our calculations confirm this mechanism for the XXZ-SSH spin chains. 

In the absence of a magnetic field, the leakage into subspaces orthogonal to the initial one (either single- or two-excitation, depending on the case of interest) exceeds $50\%$ for the waiting times considered in this work. When a magnetic field $ B_z\simeq 7 \, J/\mu_B$ is applied, this leakage drops below $0.005\%$ for all values of $\eta$ and $\Delta$, restoring the high-fidelity process observed for exchange-only spin chains. Nevertheless, the dipolar term changes the spectrum and the characteristics of the eigenstate localization. Note that although the magnetic fields strongly suppress the leakage to other subspaces, the dipolar term effect spoils the quality of the transmission.

\begin{figure}[h]
    \centering
    \includegraphics[width=0.75\textwidth]{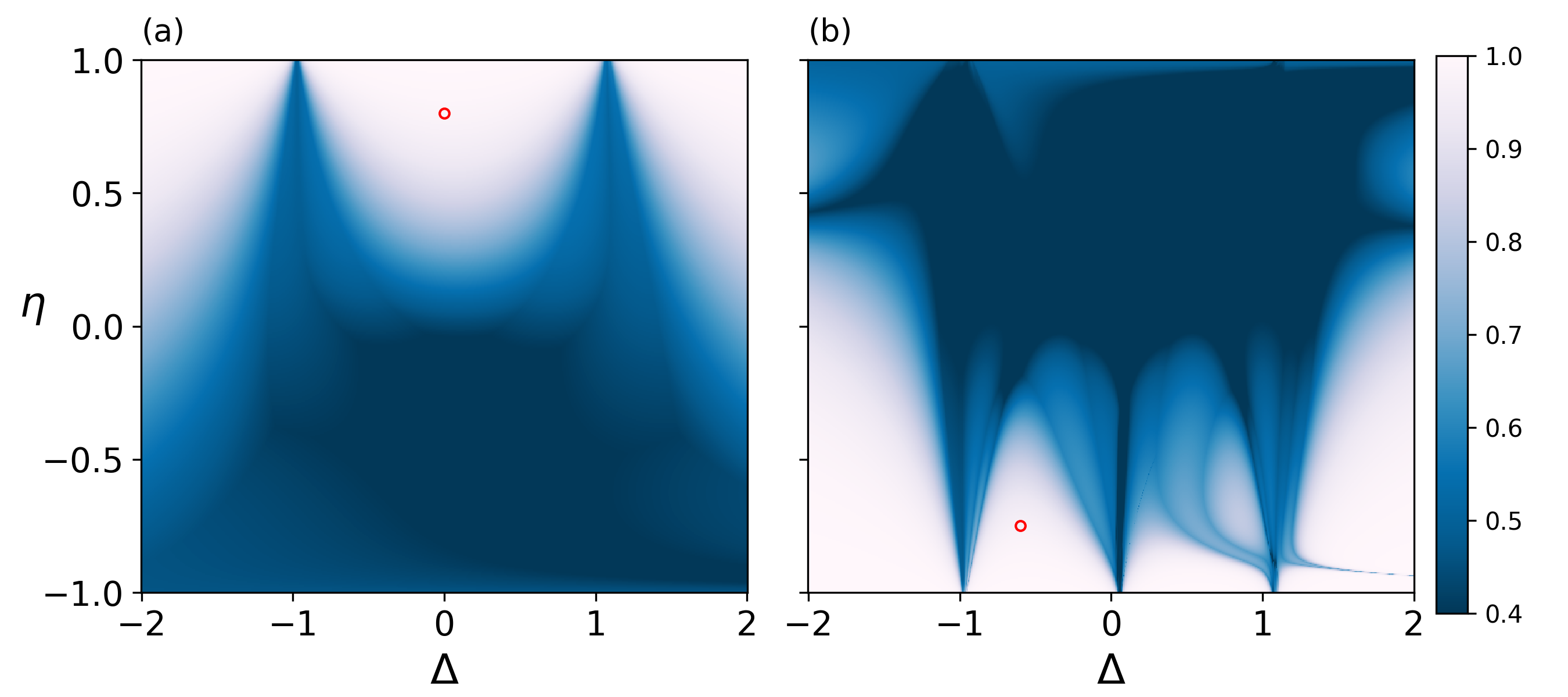}
    \includegraphics[width=0.85\textwidth]{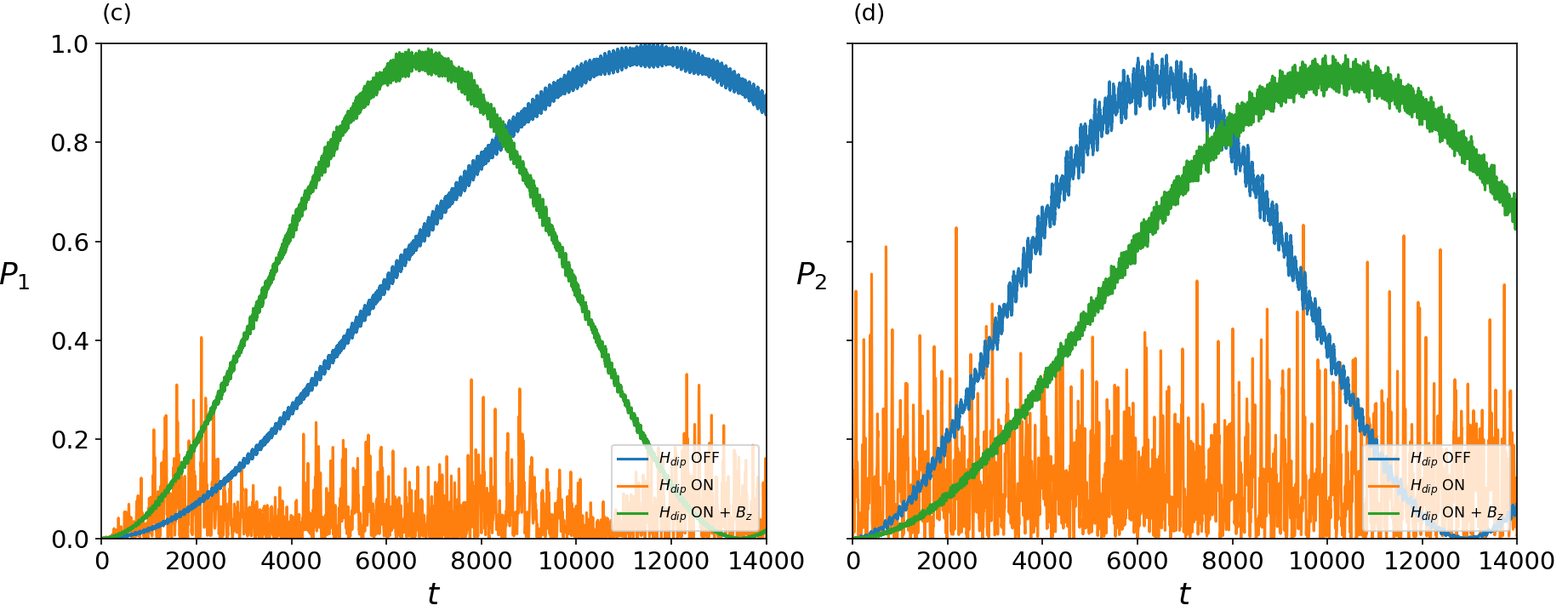}
    \caption{(a) Degree of localization for one excitation at the edges of the chain considering dipolar interaction. 
    (b) Degree of localization of two excitations at the edges of the chain considering dipolar interaction. (c) Time evolution for $\Delta=0$, $\eta=0.8$ (red point in panel (a)). (d) Time evolution for $\Delta=-0.6$, $\eta=-0.75$ (red point in panel (b)). All the calculations were done for a spin chain with $N=8$.} 
    \label{figdiploc}
\end{figure}

Before analyzing the transmission in the presence of dipolar interactions, it is essential to examine how edge-state localization behaves, both in the single-excitation and two-excitation cases. In Fig.\ref{figdiploc} (a) and (b), we plot the degree of localization $\chi$ as a function of the spin chain parameters for one and two excitations, respectively. In the single-excitation case (Figure \ref{figdiploc} (a)), localization is quantified by the total weight of the wavefunction on the first and last sites of the chain. For the two-excitation case (Figure \ref{figdiploc} (b)), we compute the weights on the first two and last two sites as in Fig. \ref{figloc2e}, capturing the degree to which the excitations remain confined near the boundaries.

Both plots are asymmetric (compare with Figs. \ref{figlocN10} (a) and \ref{figloc2e} with Figs. \ref{figdiploc} (a) and (b)); it is evident that they lack mirror symmetry around the vertical axis, $\Delta = 0$. We observe a particularly pronounced effect in the two-excitation case. Still, it is also present in the single-excitation case, where the localization appears to be shifted to the right $\delta\Delta \simeq 0.1$ when compared to the no dipolar interaction case. This behavior is reasonable, considering the complete Hamiltonian, which now has the asymmetric term $(\vec{\sigma}_i.\hat{n})(\vec{\sigma}_{j}.\hat{n})$. 

Figure~\ref{figdiploc} (c) and (d) present a good summary of the spoiling effect of the dipolar term. In both panels, the blue curve represents the transmission probability without dipolar interaction as a function of time for one or two-excitation states, shown in panels (c) and (d), respectively. The orange curves correspond to the probabilities, except that now we turn on the dipolar term. It is easy to observe the spoiling effect on the quality of the transfer process. In both cases, the probabilities fall from values near unity at their maximum to less than $0.6$. Finally, the green curves correspond to data obtained by turning on both the dipolar and Zemman terms. The Zeeman term restores the quality of the transfer. Interestingly, the maximum for the one excitation case appears earlier than when both terms are off. Conversely, for the two-excitation case, the maximum value of the probability occurs later. The delay or haste of the maximum to show up is related to the effect of the dipolar interaction on the energy gap between the border states, single and double, depending on whether the transmission is being studied for one or two excitation quantum states.


\section{Controlled Transmission of one Excitation in a Spin Chain}
\label{sroct}

In this section, we investigate the application of OCT to design time-dependent driving protocols that maximize the transmission of a single excitation from one end of the chain to the other. As the results in the previous sections make clear, obtaining high-quality transmission could take an exceedingly long transmission time.

The simpler and numerically less demanding option to control the time evolution of the quantum state of the chain starts with the application of a single time-dependent magnetic field to a single spin. In principle, the chain is controllable independently of the spin that is under the driving of the field. Numerous examples suggest that applying the field to the spin where the initial state dwells is also advisable. Options that propose applying a field, or fields, to several sites demand more computational power to calculate the driving. There are excellent reviews about the application of OCT to design control functions for controllable quantum systems in the literature \cite{Werschnik2007,Peirce1988OCT,Dong2016OCT,Hull2010OCT,James2021OCT}.  

 Besides selecting the number of control functions or the physical mechanism that couples the driving to the internal degrees of the system Hamiltonian, it is necessary to choose a target time. That time marks the moment when the task under control achieves its intended result. In our case, we expect to retrieve the information about the quantum state transferred at the target time. The selection time should not be too demanding. If the time is too short, the iterative method does not converge to a meaningful result. Let us remember that the Krotov algorithm iterates from an initial pulse shape until it reaches a minimum of the functionals constructed for this purpose. To obtain the results later shown, we set a maximum of $5000$ iterations, which is more than enough to achieve high-quality transmission probabilities over a large portion of the parameter space studied in the previous Sections. Of course, this maximum number works for short to moderate-length chains. 

Our primary objective is to understand how the effectiveness of control differs between the $\eta>0$ and $\eta<0$ dynamical regimes. With this aim, we define $T_{\text{min}}$, which represents the minimum evolution time required to design protocols to achieve high-fidelity state transfer. To study the effectiveness of the control, we focus on the probability $P_1$ and the infidelity of transmission, $1-P_1$. Note that our use of the term infidelity is different from its usual meaning. $T_{min}$ is a threshold, which is marked by a steep fall by several orders of magnitude of the infidelity value: below $T_{\text{min}}$, no control protocol can produce reliable transmission, while for longer times, successful transfer becomes possible.

\begin{figure}[h]
    \centering
    \includegraphics[width=0.75\textwidth]{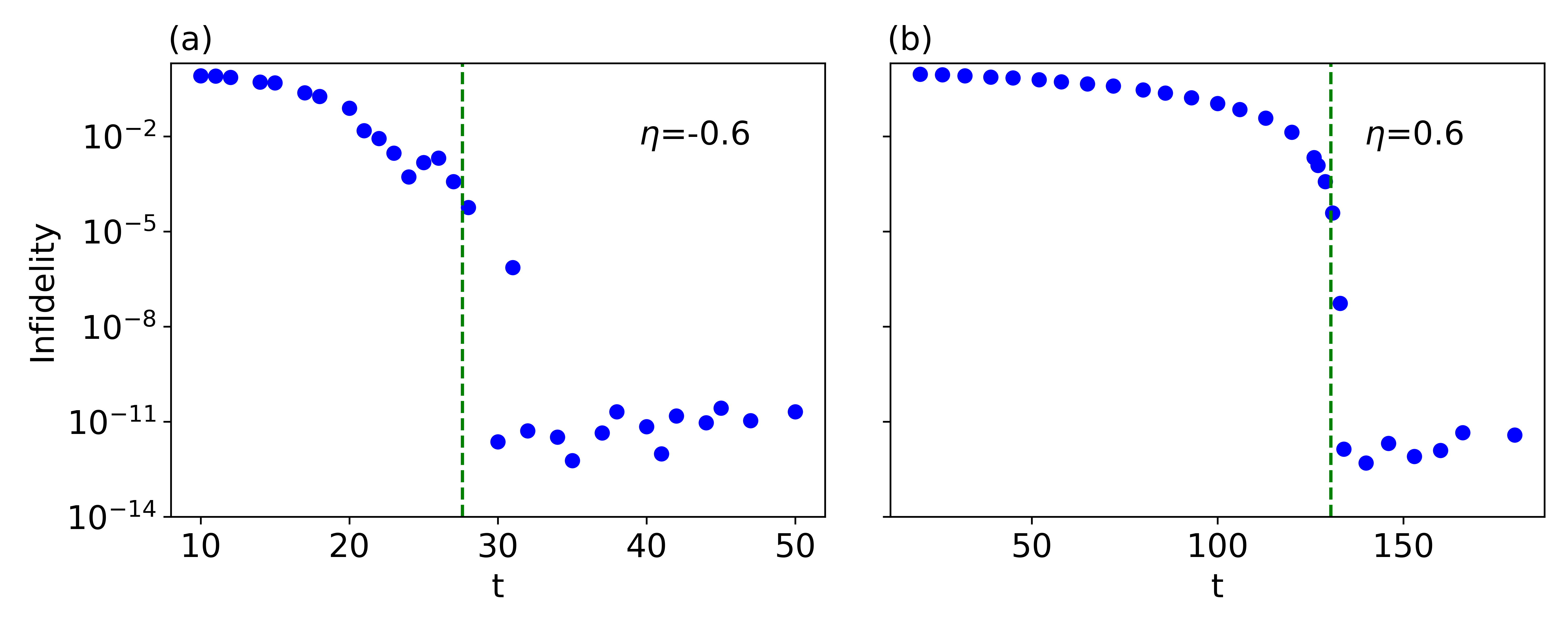}
    \vspace{0.2cm}
    \includegraphics[width=0.45\textwidth]{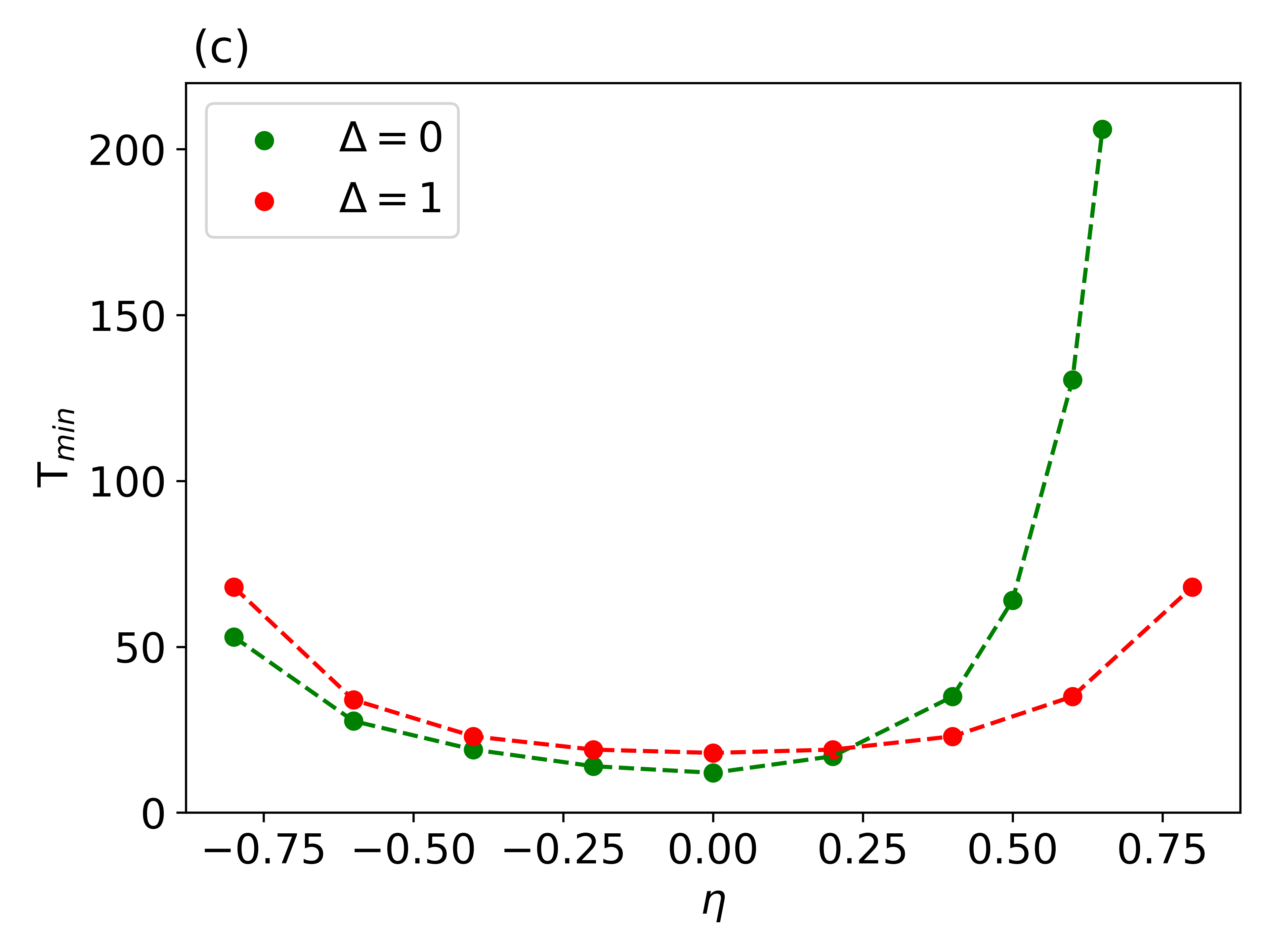}
    \caption{(a) Infidelity as a function of the time used to design the pulse for a non-topological chain. (b) Infidelity as a function of the time used to design the pulse for a topological-like chain. (c) Minimum time required to obtain an efficient pulse (red-dashed line in (a) and (b)) as a function of the dimerization parameter for an XX chain (green) and a Heisenberg chain (red).}
    \label{figoct1}
\end{figure}

The infidelity behaviour, depicted in Figs. \ref{figoct1} (a) and \ref{figoct1} (b) clearly illustrate this transition for both the non-topological and topological-like XX chains. A vertical dashed line indicates $T_{\text{min}}$, separating the controlled and poorly controlled regimes. To the left of this threshold, the optimization fails to find effective protocols, while to the right, the behavior changes qualitatively, and the optimized dynamics lead to efficient state transfer. The minimum control times vary significantly for both cases. As shown in Figs. \ref{figoct1} (a) and \ref{figoct1} (b), the minimum time required in the chain with $\eta >0$ is four times larger than in the 
non-topological case.

In Fig. \ref{figoct1} (c), we plot $T_{\text{min}}$ as a function of the dimerization parameter $\eta$ for both the XX and Heisenberg chains. This figure provides information on how topology properties and interaction type influence the control-based state transmission.
The green curve corresponds to the XX chain, while the blue one represents the dimerized Heisenberg chain. The behavior of the minimum control time in the topological regime of the XX chain is particularly noteworthy. The time required to control the transmission adequately is larger in the topological regime. For positive values of $\eta$, $T_{min}$ increases rapidly and diverges exponentially as the dimerization parameter approaches $\eta=1$. We do not observe this behavior in the Heisenberg chain, where the minimum control time remains relatively constant, regardless of whether the topological-like or non-topological properties are present. The problem with the optimization in the XX case is due to the presence of strongly localized edge states. When localization becomes strongly pronounced, finding effective control protocols becomes increasingly tricky.

We must highlight the quality of the transmission obtained through optimal control. As shown in the infidelity plots, we achieve extremely high fidelities by designing appropriate control protocols, even in this simple setup where we only use a local magnetic field to control the dynamics. This kind of control is achievable in several experimental realizations, such as atomic spin chains manipulated using ESR-STM techniques\cite{coden2024quantum,yang2017engineering,wang2025electrically,kot2023electric}. For evolution times greater than $T_{\text{min}}$, which in all cases reported in Fig. \ref{figoct1} remain below 200, the infidelity reaches values lower than $10^{-10}$. These infidelity values correspond to transmission probabilities greater than $0.9999999999$, demonstrating the possibility of achieving near-perfect state transmission by designing suitable control protocols.

Under controlled time evolution, the probability of transmission improves its value, but it is not the only quantity doing so. The arrival time also does it. For instance, using a time window $t\in \left[ 0, 2000\right]$ for autonomous evolution, we find that the maximum transmission probability achievable for $\Delta=0$, and $\eta=-0.6$ is $P_1^{(M)}=0.992$, which appears when $ t\simeq 1003.4$. The controlled time evolution achieves a probability $P_1 \gtrsim 0.9999$, for a time $T_{min}\simeq 27.5$, reducing the transmission time by a $40$ factor. For $\eta = 0.6$, we find several maxima in the time window, for instance, for $t\simeq 142$ and $t\simeq 411.7$, the local maxima are $ P_1^{(M)} \simeq 0.967$ and $P_1^{(M)} \simeq 0.975$, respectively. The controlled time evolution achieves a probability $P_1 \gtrsim 0.9999$, for a time $T_{min}\simeq 130$. It is clear that for this pair $(\eta,\Delta)$ the time needed to achieve an autonomous transmission probability larger than $0.99$ is outside the time window.


\section{Conclusions}
\label{sc}

To some extent, the fact that a Hamiltonian with only two parameters has such rich and intriguing dynamical properties is surprising. 

Our results confirm that systems with topological or similar to topological properties, $\eta>0$ and large, are more robust than non-topological ones under static disorder, as has often been stated in the literature. But, even for the short chains considered in this work, the transmission times are orders of magnitude larger than the typical transmission times found for trivial $XX$ ot $XXZ$ chains, where transmission times scale as $N$ or $N^2$.  

Another surprise comes from the comparison between the transfer ability of the system in the cases of arbitrary one and two-qubit states. Depending on the zone of the $(\Delta,\eta)$ plane chosen, the fidelity of transmission for entangled states, $F_{12}$, can reach good values close to unity for $\eta <0$ and arrival times on the order of the arrival times of the transmission of a one excitation state. There are several elongated areas near $\Delta=0$ where the transmission probability for entangled states is stable, and the tuning of the parameters of the Hamiltonian does not require great accuracy. In contrast, the quality of transmission of arbitrary two-qubits is poor, and reaching acceptable values for $F_2$  requires even larger times than those observed for similar values of $P_1$

Disorder significantly spoils efficient transmission, as demonstrated in several sections of this work. Strategies to mitigate its impact are related to both the fabrication of the devices and the characteristic times required for transmission. The topological-like regime proves to be more robust against disorder, although it typically involves long transmission times. On the other hand, in the non-topological regime, there exist parameter regions where high transmission can be achieved on shorter timescales, so the disorder does not have enough time to spoil the efficiency of the process.

As is the case for other Hamiltonians with homogeneous nearest-neighbour interactions or long-range interactions, isotropic in-plane spin interaction terms or anysotropic three-dimensional ones, the non-conservation of the total magnetization produced by the inclusion of a dipolar term and the corresponding leakage to subspaces with different magnetization than the initial quantum state, can be amended using a constant time-independent magnetic field. 

The number of points in the $(\Delta,\eta)$ plane where fast PT occurs is smaller than Figure~\ref{fign4} (b) suggests. For values of $q$ near unity, there is only a handful of solutions for Kay's equations. The lower the values of $q$, the lower the values of the transmission time. Besides, values of $\Delta,\eta$, obtained as solutions of Kay's equations with different values of $q$, are interspersed with each other, so chains with similar $\Delta,\eta$ parameters can present PT at very different transmission times. That the points of the $\Delta,\eta$ plane compatible with PT and very different arrival times are interspersed has several consequences. Each point on the parameter plane where PT is possible has a zone surrounding it,  and where the fulfillment of Kay's conditions is only approximate. It is well-known that under this condition, the exact periodicity of PT changes, and the time evolution becomes quasi-periodic. The quasi-periodicity is characteristic of the pretty good quantum state transfer scenario. It is clear that very close to the parameters where PT occurs, the $q$ values needed by Kay's equations determine the characteristics of the quasi-periodic time evolution. As a consequence, there must be a crossover in the time evolution characteristics from one set of parameters compatible with PT to another set. 

The ability of the OCT method to find pulses whose forcing produces faster transmission with a larger transmission probability than the natural evolution is remarkable. Nevertheless, testing its scaling while increasing the length of the chains under study is a daunting prospect when $\eta$ is positive and larger than $0.6$. The failure of OCT to find a fast and successful control strategy does not mean that there are no other control methods that could be more successful at this particular task. 


\
\bibliography{biblio}{}

\begin{thebibliography}{49}%
\makeatletter
\providecommand \@ifxundefined [1]{%
 \@ifx{#1\undefined}
}%
\providecommand \@ifnum [1]{%
 \ifnum #1\expandafter \@firstoftwo
 \else \expandafter \@secondoftwo
 \fi
}%
\providecommand \@ifx [1]{%
 \ifx #1\expandafter \@firstoftwo
 \else \expandafter \@secondoftwo
 \fi
}%
\providecommand \natexlab [1]{#1}%
\providecommand \enquote  [1]{``#1''}%
\providecommand \bibnamefont  [1]{#1}%
\providecommand \bibfnamefont [1]{#1}%
\providecommand \citenamefont [1]{#1}%
\providecommand \href@noop [0]{\@secondoftwo}%
\providecommand \href [0]{\begingroup \@sanitize@url \@href}%
\providecommand \@href[1]{\@@startlink{#1}\@@href}%
\providecommand \@@href[1]{\endgroup#1\@@endlink}%
\providecommand \@sanitize@url [0]{\catcode `\\12\catcode `\$12\catcode `\&12\catcode `\#12\catcode `\^12\catcode `\_12\catcode `\%12\relax}%
\providecommand \@@startlink[1]{}%
\providecommand \@@endlink[0]{}%
\providecommand \url  [0]{\begingroup\@sanitize@url \@url }%
\providecommand \@url [1]{\endgroup\@href {#1}{\urlprefix }}%
\providecommand \urlprefix  [0]{URL }%
\providecommand \Eprint [0]{\href }%
\providecommand \doibase [0]{http://dx.doi.org/}%
\providecommand \selectlanguage [0]{\@gobble}%
\providecommand \bibinfo  [0]{\@secondoftwo}%
\providecommand \bibfield  [0]{\@secondoftwo}%
\providecommand \translation [1]{[#1]}%
\providecommand \BibitemOpen [0]{}%
\providecommand \bibitemStop [0]{}%
\providecommand \bibitemNoStop [0]{.\EOS\space}%
\providecommand \EOS [0]{\spacefactor3000\relax}%
\providecommand \BibitemShut  [1]{\csname bibitem#1\endcsname}%
\let\auto@bib@innerbib\@empty
\bibitem [{\citenamefont {Xie}\ \emph {et~al.}(2023)\citenamefont {Xie}, \citenamefont {Kay},\ and\ \citenamefont {Tamon}}]{Xie2023}%
  \BibitemOpen
  \bibfield  {author} {\bibinfo {author} {\bibfnamefont {W.}~\bibnamefont {Xie}}, \bibinfo {author} {\bibfnamefont {A.}~\bibnamefont {Kay}}, \ and\ \bibinfo {author} {\bibfnamefont {C.}~\bibnamefont {Tamon}},\ }\href {\doibase 10.1103/PhysRevA.108.012408} {\bibfield  {journal} {\bibinfo  {journal} {Phys. Rev. A}\ }\textbf {\bibinfo {volume} {108}},\ \bibinfo {pages} {012408} (\bibinfo {year} {2023})}\BibitemShut {NoStop}%
\bibitem [{\citenamefont {Maleki}\ and\ \citenamefont {Zheltikov}(2021)}]{Maleki2021}%
  \BibitemOpen
  \bibfield  {author} {\bibinfo {author} {\bibfnamefont {Y.}~\bibnamefont {Maleki}}\ and\ \bibinfo {author} {\bibfnamefont {A.~M.}\ \bibnamefont {Zheltikov}},\ }\href {\doibase https://doi.org/10.1016/j.optcom.2021.126870} {\bibfield  {journal} {\bibinfo  {journal} {Optics Communications}\ }\textbf {\bibinfo {volume} {496}},\ \bibinfo {pages} {126870} (\bibinfo {year} {2021})}\BibitemShut {NoStop}%
\bibitem [{\citenamefont {Stehlik}\ \emph {et~al.}(2021)\citenamefont {Stehlik}, \citenamefont {Zajac}, \citenamefont {Underwood}, \citenamefont {Phung}, \citenamefont {Blair}, \citenamefont {Carnevale}, \citenamefont {Klaus}, \citenamefont {Keefe}, \citenamefont {Carniol}, \citenamefont {Kumph}, \citenamefont {Steffen},\ and\ \citenamefont {Dial}}]{Stehlik2021}%
  \BibitemOpen
  \bibfield  {author} {\bibinfo {author} {\bibfnamefont {J.}~\bibnamefont {Stehlik}}, \bibinfo {author} {\bibfnamefont {D.~M.}\ \bibnamefont {Zajac}}, \bibinfo {author} {\bibfnamefont {D.~L.}\ \bibnamefont {Underwood}}, \bibinfo {author} {\bibfnamefont {T.}~\bibnamefont {Phung}}, \bibinfo {author} {\bibfnamefont {J.}~\bibnamefont {Blair}}, \bibinfo {author} {\bibfnamefont {S.}~\bibnamefont {Carnevale}}, \bibinfo {author} {\bibfnamefont {D.}~\bibnamefont {Klaus}}, \bibinfo {author} {\bibfnamefont {G.~A.}\ \bibnamefont {Keefe}}, \bibinfo {author} {\bibfnamefont {A.}~\bibnamefont {Carniol}}, \bibinfo {author} {\bibfnamefont {M.}~\bibnamefont {Kumph}}, \bibinfo {author} {\bibfnamefont {M.}~\bibnamefont {Steffen}}, \ and\ \bibinfo {author} {\bibfnamefont {O.~E.}\ \bibnamefont {Dial}},\ }\href {\doibase 10.1103/PhysRevLett.127.080505} {\bibfield  {journal} {\bibinfo  {journal} {Phys. Rev. Lett.}\ }\textbf {\bibinfo {volume} {127}},\ \bibinfo {pages} {080505} (\bibinfo {year} {2021})}\BibitemShut {NoStop}%
\bibitem [{\citenamefont {Serra}\ \emph {et~al.}(2025)\citenamefont {Serra}, \citenamefont {Ferrón},\ and\ \citenamefont {Osenda}}]{serra2025perfect}%
  \BibitemOpen
  \bibfield  {author} {\bibinfo {author} {\bibfnamefont {P.}~\bibnamefont {Serra}}, \bibinfo {author} {\bibfnamefont {A.}~\bibnamefont {Ferrón}}, \ and\ \bibinfo {author} {\bibfnamefont {O.}~\bibnamefont {Osenda}},\ }\href {\doibase https://doi.org/10.1016/j.physleta.2025.130817} {\bibfield  {journal} {\bibinfo  {journal} {Physics Letters A}\ }\textbf {\bibinfo {volume} {556}},\ \bibinfo {pages} {130817} (\bibinfo {year} {2025})}\BibitemShut {NoStop}%
\bibitem [{\citenamefont {Li}\ \emph {et~al.}(2018)\citenamefont {Li}, \citenamefont {Ma}, \citenamefont {Han}, \citenamefont {Chen}, \citenamefont {Xu}, \citenamefont {Cai}, \citenamefont {Wang}, \citenamefont {Song}, \citenamefont {Xue}, \citenamefont {Yin},\ and\ \citenamefont {Sun}}]{Li2018transmon}%
  \BibitemOpen
  \bibfield  {author} {\bibinfo {author} {\bibfnamefont {X.}~\bibnamefont {Li}}, \bibinfo {author} {\bibfnamefont {Y.}~\bibnamefont {Ma}}, \bibinfo {author} {\bibfnamefont {J.}~\bibnamefont {Han}}, \bibinfo {author} {\bibfnamefont {T.}~\bibnamefont {Chen}}, \bibinfo {author} {\bibfnamefont {Y.}~\bibnamefont {Xu}}, \bibinfo {author} {\bibfnamefont {W.}~\bibnamefont {Cai}}, \bibinfo {author} {\bibfnamefont {H.}~\bibnamefont {Wang}}, \bibinfo {author} {\bibfnamefont {Y.}~\bibnamefont {Song}}, \bibinfo {author} {\bibfnamefont {Z.-Y.}\ \bibnamefont {Xue}}, \bibinfo {author} {\bibfnamefont {Z.-q.}\ \bibnamefont {Yin}}, \ and\ \bibinfo {author} {\bibfnamefont {L.}~\bibnamefont {Sun}},\ }\href {\doibase 10.1103/PhysRevApplied.10.054009} {\bibfield  {journal} {\bibinfo  {journal} {Phys. Rev. Appl.}\ }\textbf {\bibinfo {volume} {10}},\ \bibinfo {pages} {054009} (\bibinfo {year} {2018})}\BibitemShut {NoStop}%
\bibitem [{\citenamefont {Su}\ \emph {et~al.}(1979)\citenamefont {Su}, \citenamefont {Schrieffer},\ and\ \citenamefont {Heeger}}]{Su1979}%
  \BibitemOpen
  \bibfield  {author} {\bibinfo {author} {\bibfnamefont {W.~P.}\ \bibnamefont {Su}}, \bibinfo {author} {\bibfnamefont {J.~R.}\ \bibnamefont {Schrieffer}}, \ and\ \bibinfo {author} {\bibfnamefont {A.~J.}\ \bibnamefont {Heeger}},\ }\href {\doibase 10.1103/PhysRevLett.42.1698} {\bibfield  {journal} {\bibinfo  {journal} {Phys. Rev. Lett.}\ }\textbf {\bibinfo {volume} {42}},\ \bibinfo {pages} {1698} (\bibinfo {year} {1979})}\BibitemShut {NoStop}%
\bibitem [{\citenamefont {Xie}\ \emph {et~al.}(2019)\citenamefont {Xie}, \citenamefont {Gou}, \citenamefont {Xiao}, \citenamefont {Gadway},\ and\ \citenamefont {Yan}}]{Xie2019}%
  \BibitemOpen
  \bibfield  {author} {\bibinfo {author} {\bibfnamefont {D.}~\bibnamefont {Xie}}, \bibinfo {author} {\bibfnamefont {W.}~\bibnamefont {Gou}}, \bibinfo {author} {\bibfnamefont {T.}~\bibnamefont {Xiao}}, \bibinfo {author} {\bibfnamefont {B.}~\bibnamefont {Gadway}}, \ and\ \bibinfo {author} {\bibfnamefont {B.}~\bibnamefont {Yan}},\ }\href {\doibase 10.1038/s41534-019-0159-6} {\bibfield  {journal} {\bibinfo  {journal} {npj Quantum Information}\ }\textbf {\bibinfo {volume} {5}},\ \bibinfo {pages} {55} (\bibinfo {year} {2019})}\BibitemShut {NoStop}%
\bibitem [{\citenamefont {Estarellas}\ \emph {et~al.}(2017)\citenamefont {Estarellas}, \citenamefont {D'Amico},\ and\ \citenamefont {Spiller}}]{Estarellas2017}%
  \BibitemOpen
  \bibfield  {author} {\bibinfo {author} {\bibfnamefont {M.~P.}\ \bibnamefont {Estarellas}}, \bibinfo {author} {\bibfnamefont {I.}~\bibnamefont {D'Amico}}, \ and\ \bibinfo {author} {\bibfnamefont {T.~P.}\ \bibnamefont {Spiller}},\ }\href {\doibase 10.1038/srep42904} {\bibfield  {journal} {\bibinfo  {journal} {Scientific Reports}\ }\textbf {\bibinfo {volume} {7}},\ \bibinfo {pages} {42904} (\bibinfo {year} {2017})}\BibitemShut {NoStop}%
\bibitem [{\citenamefont {Wang}\ \emph {et~al.}(2022)\citenamefont {Wang}, \citenamefont {Li}, \citenamefont {Gong},\ and\ \citenamefont {Liu}}]{Wang2022}%
  \BibitemOpen
  \bibfield  {author} {\bibinfo {author} {\bibfnamefont {C.}~\bibnamefont {Wang}}, \bibinfo {author} {\bibfnamefont {L.}~\bibnamefont {Li}}, \bibinfo {author} {\bibfnamefont {J.}~\bibnamefont {Gong}}, \ and\ \bibinfo {author} {\bibfnamefont {Y.-x.}\ \bibnamefont {Liu}},\ }\href {\doibase 10.1103/PhysRevA.106.052411} {\bibfield  {journal} {\bibinfo  {journal} {Phys. Rev. A}\ }\textbf {\bibinfo {volume} {106}},\ \bibinfo {pages} {052411} (\bibinfo {year} {2022})}\BibitemShut {NoStop}%
\bibitem [{\citenamefont {Qi}\ \emph {et~al.}(2024)\citenamefont {Qi}, \citenamefont {Li}, \citenamefont {Han}, \citenamefont {Li}, \citenamefont {Zhang},\ and\ \citenamefont {He}}]{Qi2024}%
  \BibitemOpen
  \bibfield  {author} {\bibinfo {author} {\bibfnamefont {L.}~\bibnamefont {Qi}}, \bibinfo {author} {\bibfnamefont {Q.-N.}\ \bibnamefont {Li}}, \bibinfo {author} {\bibfnamefont {N.}~\bibnamefont {Han}}, \bibinfo {author} {\bibfnamefont {M.}~\bibnamefont {Li}}, \bibinfo {author} {\bibfnamefont {X.-Y.}\ \bibnamefont {Zhang}}, \ and\ \bibinfo {author} {\bibfnamefont {A.-L.}\ \bibnamefont {He}},\ }\href {\doibase 10.1103/PhysRevA.109.032428} {\bibfield  {journal} {\bibinfo  {journal} {Phys. Rev. A}\ }\textbf {\bibinfo {volume} {109}},\ \bibinfo {pages} {032428} (\bibinfo {year} {2024})}\BibitemShut {NoStop}%
\bibitem [{\citenamefont {Legg}(2025)}]{legg2025commentinasalhybriddevices}%
  \BibitemOpen
  \bibfield  {author} {\bibinfo {author} {\bibfnamefont {H.~F.}\ \bibnamefont {Legg}},\ }\href {https://arxiv.org/abs/2502.19560} {\enquote {\bibinfo {title} {Comment on "inas-al hybrid devices passing the topological gap protocol", microsoft quantum, phys. rev. b 107, 245423 (2023)},}\ } (\bibinfo {year} {2025}),\ \Eprint {http://arxiv.org/abs/2502.19560} {arXiv:2502.19560 [cond-mat.mes-hall]} \BibitemShut {NoStop}%
\bibitem [{\citenamefont {Aghaee}\ \emph {et~al.}(2023)\citenamefont {Aghaee}, \citenamefont {Akkala}, \citenamefont {Alam}, \citenamefont {Ali}, \citenamefont {Alcaraz~Ramirez}, \citenamefont {Andrzejczuk}, \citenamefont {Antipov}, \citenamefont {Aseev}, \citenamefont {Astafev}, \citenamefont {Bauer}, \citenamefont {Becker}, \citenamefont {Boddapati}, \citenamefont {Boekhout}, \citenamefont {Bommer}, \citenamefont {Bosma}, \citenamefont {Bourdet}, \citenamefont {Boutin}, \citenamefont {Caroff}, \citenamefont {Casparis}, \citenamefont {Cassidy}, \citenamefont {Chatoor}, \citenamefont {Christensen}, \citenamefont {Clay}, \citenamefont {Cole}, \citenamefont {Corsetti}, \citenamefont {Cui}, \citenamefont {Dalampiras}, \citenamefont {Dokania}, \citenamefont {de~Lange}, \citenamefont {de~Moor}, \citenamefont {Estrada Salda\~na}, \citenamefont {Fallahi}, \citenamefont {Fathabad}, \citenamefont {Gamble}, \citenamefont {Gardner}, \citenamefont {Govender}, \citenamefont {Griggio}, \citenamefont {Grigoryan},
  \citenamefont {Gronin}, \citenamefont {Gukelberger}, \citenamefont {Hansen}, \citenamefont {Heedt}, \citenamefont {Herranz~Zamorano}, \citenamefont {Ho}, \citenamefont {Holgaard}, \citenamefont {Ingerslev}, \citenamefont {Johansson}, \citenamefont {Jones}, \citenamefont {Kallaher}, \citenamefont {Karimi}, \citenamefont {Karzig}, \citenamefont {King}, \citenamefont {Kloster}, \citenamefont {Knapp}, \citenamefont {Kocon}, \citenamefont {Koski}, \citenamefont {Kostamo}, \citenamefont {Krogstrup}, \citenamefont {Kumar}, \citenamefont {Laeven}, \citenamefont {Larsen}, \citenamefont {Li}, \citenamefont {Lindemann}, \citenamefont {Love}, \citenamefont {Lutchyn}, \citenamefont {Madsen}, \citenamefont {Manfra}, \citenamefont {Markussen}, \citenamefont {Martinez}, \citenamefont {McNeil}, \citenamefont {Memisevic}, \citenamefont {Morgan}, \citenamefont {Mullally}, \citenamefont {Nayak}, \citenamefont {Nielsen}, \citenamefont {Nielsen}, \citenamefont {Nijholt}, \citenamefont {Nurmohamed}, \citenamefont {O'Farrell},
  \citenamefont {Otani}, \citenamefont {Pauka}, \citenamefont {Petersson}, \citenamefont {Petit}, \citenamefont {Pikulin}, \citenamefont {Preiss}, \citenamefont {Quintero-Perez}, \citenamefont {Rajpalke}, \citenamefont {Rasmussen}, \citenamefont {Razmadze}, \citenamefont {Reentila}, \citenamefont {Reilly}, \citenamefont {Rouse}, \citenamefont {Sadovskyy}, \citenamefont {Sainiemi}, \citenamefont {Schreppler}, \citenamefont {Sidorkin}, \citenamefont {Singh}, \citenamefont {Singh}, \citenamefont {Sinha}, \citenamefont {Sohr}, \citenamefont {Stankevi\ifmmode~\check{c}\else \v{c}\fi{}}, \citenamefont {Stek}, \citenamefont {Suominen}, \citenamefont {Suter}, \citenamefont {Svidenko}, \citenamefont {Teicher}, \citenamefont {Temuerhan}, \citenamefont {Thiyagarajah}, \citenamefont {Tholapi}, \citenamefont {Thomas}, \citenamefont {Toomey}, \citenamefont {Upadhyay}, \citenamefont {Urban}, \citenamefont {Vaitiek\ifmmode~\dot{e}\else \.{e}\fi{}nas}, \citenamefont {Van~Hoogdalem}, \citenamefont {Van~Woerkom}, \citenamefont
  {Viazmitinov}, \citenamefont {Vogel}, \citenamefont {Waddy}, \citenamefont {Watson}, \citenamefont {Weston}, \citenamefont {Winkler}, \citenamefont {Yang}, \citenamefont {Yau}, \citenamefont {Yi}, \citenamefont {Yucelen}, \citenamefont {Webster}, \citenamefont {Zeisel},\ and\ \citenamefont {Zhao}}]{Aghaee2023hybridtopo}%
  \BibitemOpen
  \bibfield  {author} {\bibinfo {author} {\bibfnamefont {M.}~\bibnamefont {Aghaee}}, \bibinfo {author} {\bibfnamefont {A.}~\bibnamefont {Akkala}}, \bibinfo {author} {\bibfnamefont {Z.}~\bibnamefont {Alam}}, \bibinfo {author} {\bibfnamefont {R.}~\bibnamefont {Ali}}, \bibinfo {author} {\bibfnamefont {A.}~\bibnamefont {Alcaraz~Ramirez}}, \bibinfo {author} {\bibfnamefont {M.}~\bibnamefont {Andrzejczuk}}, \bibinfo {author} {\bibfnamefont {A.~E.}\ \bibnamefont {Antipov}}, \bibinfo {author} {\bibfnamefont {P.}~\bibnamefont {Aseev}}, \bibinfo {author} {\bibfnamefont {M.}~\bibnamefont {Astafev}}, \bibinfo {author} {\bibfnamefont {B.}~\bibnamefont {Bauer}}, \bibinfo {author} {\bibfnamefont {J.}~\bibnamefont {Becker}}, \bibinfo {author} {\bibfnamefont {S.}~\bibnamefont {Boddapati}}, \bibinfo {author} {\bibfnamefont {F.}~\bibnamefont {Boekhout}}, \bibinfo {author} {\bibfnamefont {J.}~\bibnamefont {Bommer}}, \bibinfo {author} {\bibfnamefont {T.}~\bibnamefont {Bosma}}, \bibinfo {author} {\bibfnamefont {L.}~\bibnamefont
  {Bourdet}}, \bibinfo {author} {\bibfnamefont {S.}~\bibnamefont {Boutin}}, \bibinfo {author} {\bibfnamefont {P.}~\bibnamefont {Caroff}}, \bibinfo {author} {\bibfnamefont {L.}~\bibnamefont {Casparis}}, \bibinfo {author} {\bibfnamefont {M.}~\bibnamefont {Cassidy}}, \bibinfo {author} {\bibfnamefont {S.}~\bibnamefont {Chatoor}}, \bibinfo {author} {\bibfnamefont {A.~W.}\ \bibnamefont {Christensen}}, \bibinfo {author} {\bibfnamefont {N.}~\bibnamefont {Clay}}, \bibinfo {author} {\bibfnamefont {W.~S.}\ \bibnamefont {Cole}}, \bibinfo {author} {\bibfnamefont {F.}~\bibnamefont {Corsetti}}, \bibinfo {author} {\bibfnamefont {A.}~\bibnamefont {Cui}}, \bibinfo {author} {\bibfnamefont {P.}~\bibnamefont {Dalampiras}}, \bibinfo {author} {\bibfnamefont {A.}~\bibnamefont {Dokania}}, \bibinfo {author} {\bibfnamefont {G.}~\bibnamefont {de~Lange}}, \bibinfo {author} {\bibfnamefont {M.}~\bibnamefont {de~Moor}}, \bibinfo {author} {\bibfnamefont {J.~C.}\ \bibnamefont {Estrada Salda\~na}}, \bibinfo {author} {\bibfnamefont
  {S.}~\bibnamefont {Fallahi}}, \bibinfo {author} {\bibfnamefont {Z.~H.}\ \bibnamefont {Fathabad}}, \bibinfo {author} {\bibfnamefont {J.}~\bibnamefont {Gamble}}, \bibinfo {author} {\bibfnamefont {G.}~\bibnamefont {Gardner}}, \bibinfo {author} {\bibfnamefont {D.}~\bibnamefont {Govender}}, \bibinfo {author} {\bibfnamefont {F.}~\bibnamefont {Griggio}}, \bibinfo {author} {\bibfnamefont {R.}~\bibnamefont {Grigoryan}}, \bibinfo {author} {\bibfnamefont {S.}~\bibnamefont {Gronin}}, \bibinfo {author} {\bibfnamefont {J.}~\bibnamefont {Gukelberger}}, \bibinfo {author} {\bibfnamefont {E.~B.}\ \bibnamefont {Hansen}}, \bibinfo {author} {\bibfnamefont {S.}~\bibnamefont {Heedt}}, \bibinfo {author} {\bibfnamefont {J.}~\bibnamefont {Herranz~Zamorano}}, \bibinfo {author} {\bibfnamefont {S.}~\bibnamefont {Ho}}, \bibinfo {author} {\bibfnamefont {U.~L.}\ \bibnamefont {Holgaard}}, \bibinfo {author} {\bibfnamefont {H.}~\bibnamefont {Ingerslev}}, \bibinfo {author} {\bibfnamefont {L.}~\bibnamefont {Johansson}}, \bibinfo {author}
  {\bibfnamefont {J.}~\bibnamefont {Jones}}, \bibinfo {author} {\bibfnamefont {R.}~\bibnamefont {Kallaher}}, \bibinfo {author} {\bibfnamefont {F.}~\bibnamefont {Karimi}}, \bibinfo {author} {\bibfnamefont {T.}~\bibnamefont {Karzig}}, \bibinfo {author} {\bibfnamefont {E.}~\bibnamefont {King}}, \bibinfo {author} {\bibfnamefont {M.~E.}\ \bibnamefont {Kloster}}, \bibinfo {author} {\bibfnamefont {C.}~\bibnamefont {Knapp}}, \bibinfo {author} {\bibfnamefont {D.}~\bibnamefont {Kocon}}, \bibinfo {author} {\bibfnamefont {J.}~\bibnamefont {Koski}}, \bibinfo {author} {\bibfnamefont {P.}~\bibnamefont {Kostamo}}, \bibinfo {author} {\bibfnamefont {P.}~\bibnamefont {Krogstrup}}, \bibinfo {author} {\bibfnamefont {M.}~\bibnamefont {Kumar}}, \bibinfo {author} {\bibfnamefont {T.}~\bibnamefont {Laeven}}, \bibinfo {author} {\bibfnamefont {T.}~\bibnamefont {Larsen}}, \bibinfo {author} {\bibfnamefont {K.}~\bibnamefont {Li}}, \bibinfo {author} {\bibfnamefont {T.}~\bibnamefont {Lindemann}}, \bibinfo {author} {\bibfnamefont
  {J.}~\bibnamefont {Love}}, \bibinfo {author} {\bibfnamefont {R.}~\bibnamefont {Lutchyn}}, \bibinfo {author} {\bibfnamefont {M.~H.}\ \bibnamefont {Madsen}}, \bibinfo {author} {\bibfnamefont {M.}~\bibnamefont {Manfra}}, \bibinfo {author} {\bibfnamefont {S.}~\bibnamefont {Markussen}}, \bibinfo {author} {\bibfnamefont {E.}~\bibnamefont {Martinez}}, \bibinfo {author} {\bibfnamefont {R.}~\bibnamefont {McNeil}}, \bibinfo {author} {\bibfnamefont {E.}~\bibnamefont {Memisevic}}, \bibinfo {author} {\bibfnamefont {T.}~\bibnamefont {Morgan}}, \bibinfo {author} {\bibfnamefont {A.}~\bibnamefont {Mullally}}, \bibinfo {author} {\bibfnamefont {C.}~\bibnamefont {Nayak}}, \bibinfo {author} {\bibfnamefont {J.}~\bibnamefont {Nielsen}}, \bibinfo {author} {\bibfnamefont {W.~H.~P.}\ \bibnamefont {Nielsen}}, \bibinfo {author} {\bibfnamefont {B.}~\bibnamefont {Nijholt}}, \bibinfo {author} {\bibfnamefont {A.}~\bibnamefont {Nurmohamed}}, \bibinfo {author} {\bibfnamefont {E.}~\bibnamefont {O'Farrell}}, \bibinfo {author} {\bibfnamefont
  {K.}~\bibnamefont {Otani}}, \bibinfo {author} {\bibfnamefont {S.}~\bibnamefont {Pauka}}, \bibinfo {author} {\bibfnamefont {K.}~\bibnamefont {Petersson}}, \bibinfo {author} {\bibfnamefont {L.}~\bibnamefont {Petit}}, \bibinfo {author} {\bibfnamefont {D.~I.}\ \bibnamefont {Pikulin}}, \bibinfo {author} {\bibfnamefont {F.}~\bibnamefont {Preiss}}, \bibinfo {author} {\bibfnamefont {M.}~\bibnamefont {Quintero-Perez}}, \bibinfo {author} {\bibfnamefont {M.}~\bibnamefont {Rajpalke}}, \bibinfo {author} {\bibfnamefont {K.}~\bibnamefont {Rasmussen}}, \bibinfo {author} {\bibfnamefont {D.}~\bibnamefont {Razmadze}}, \bibinfo {author} {\bibfnamefont {O.}~\bibnamefont {Reentila}}, \bibinfo {author} {\bibfnamefont {D.}~\bibnamefont {Reilly}}, \bibinfo {author} {\bibfnamefont {R.}~\bibnamefont {Rouse}}, \bibinfo {author} {\bibfnamefont {I.}~\bibnamefont {Sadovskyy}}, \bibinfo {author} {\bibfnamefont {L.}~\bibnamefont {Sainiemi}}, \bibinfo {author} {\bibfnamefont {S.}~\bibnamefont {Schreppler}}, \bibinfo {author} {\bibfnamefont
  {V.}~\bibnamefont {Sidorkin}}, \bibinfo {author} {\bibfnamefont {A.}~\bibnamefont {Singh}}, \bibinfo {author} {\bibfnamefont {S.}~\bibnamefont {Singh}}, \bibinfo {author} {\bibfnamefont {S.}~\bibnamefont {Sinha}}, \bibinfo {author} {\bibfnamefont {P.}~\bibnamefont {Sohr}}, \bibinfo {author} {\bibfnamefont {T.~c.~v.}\ \bibnamefont {Stankevi\ifmmode~\check{c}\else \v{c}\fi{}}}, \bibinfo {author} {\bibfnamefont {L.}~\bibnamefont {Stek}}, \bibinfo {author} {\bibfnamefont {H.}~\bibnamefont {Suominen}}, \bibinfo {author} {\bibfnamefont {J.}~\bibnamefont {Suter}}, \bibinfo {author} {\bibfnamefont {V.}~\bibnamefont {Svidenko}}, \bibinfo {author} {\bibfnamefont {S.}~\bibnamefont {Teicher}}, \bibinfo {author} {\bibfnamefont {M.}~\bibnamefont {Temuerhan}}, \bibinfo {author} {\bibfnamefont {N.}~\bibnamefont {Thiyagarajah}}, \bibinfo {author} {\bibfnamefont {R.}~\bibnamefont {Tholapi}}, \bibinfo {author} {\bibfnamefont {M.}~\bibnamefont {Thomas}}, \bibinfo {author} {\bibfnamefont {E.}~\bibnamefont {Toomey}}, \bibinfo
  {author} {\bibfnamefont {S.}~\bibnamefont {Upadhyay}}, \bibinfo {author} {\bibfnamefont {I.}~\bibnamefont {Urban}}, \bibinfo {author} {\bibfnamefont {S.}~\bibnamefont {Vaitiek\ifmmode~\dot{e}\else \.{e}\fi{}nas}}, \bibinfo {author} {\bibfnamefont {K.}~\bibnamefont {Van~Hoogdalem}}, \bibinfo {author} {\bibfnamefont {D.}~\bibnamefont {Van~Woerkom}}, \bibinfo {author} {\bibfnamefont {D.~V.}\ \bibnamefont {Viazmitinov}}, \bibinfo {author} {\bibfnamefont {D.}~\bibnamefont {Vogel}}, \bibinfo {author} {\bibfnamefont {S.}~\bibnamefont {Waddy}}, \bibinfo {author} {\bibfnamefont {J.}~\bibnamefont {Watson}}, \bibinfo {author} {\bibfnamefont {J.}~\bibnamefont {Weston}}, \bibinfo {author} {\bibfnamefont {G.~W.}\ \bibnamefont {Winkler}}, \bibinfo {author} {\bibfnamefont {C.~K.}\ \bibnamefont {Yang}}, \bibinfo {author} {\bibfnamefont {S.}~\bibnamefont {Yau}}, \bibinfo {author} {\bibfnamefont {D.}~\bibnamefont {Yi}}, \bibinfo {author} {\bibfnamefont {E.}~\bibnamefont {Yucelen}}, \bibinfo {author} {\bibfnamefont
  {A.}~\bibnamefont {Webster}}, \bibinfo {author} {\bibfnamefont {R.}~\bibnamefont {Zeisel}}, \ and\ \bibinfo {author} {\bibfnamefont {R.}~\bibnamefont {Zhao}} (\bibinfo {collaboration} {Microsoft Quantum}),\ }\href {\doibase 10.1103/PhysRevB.107.245423} {\bibfield  {journal} {\bibinfo  {journal} {Phys. Rev. B}\ }\textbf {\bibinfo {volume} {107}},\ \bibinfo {pages} {245423} (\bibinfo {year} {2023})}\BibitemShut {NoStop}%
\bibitem [{\citenamefont {Wang}\ \emph {et~al.}(2016)\citenamefont {Wang}, \citenamefont {Burgarth},\ and\ \citenamefont {Schirmer}}]{Wang2016}%
  \BibitemOpen
  \bibfield  {author} {\bibinfo {author} {\bibfnamefont {X.}~\bibnamefont {Wang}}, \bibinfo {author} {\bibfnamefont {D.}~\bibnamefont {Burgarth}}, \ and\ \bibinfo {author} {\bibfnamefont {S.}~\bibnamefont {Schirmer}},\ }\href {\doibase 10.1103/PhysRevA.94.052319} {\bibfield  {journal} {\bibinfo  {journal} {Phys. Rev. A}\ }\textbf {\bibinfo {volume} {94}},\ \bibinfo {pages} {052319} (\bibinfo {year} {2016})}\BibitemShut {NoStop}%
\bibitem [{\citenamefont {Burgarth}\ \emph {et~al.}(2010)\citenamefont {Burgarth}, \citenamefont {Maruyama}, \citenamefont {Murphy}, \citenamefont {Montangero}, \citenamefont {Calarco}, \citenamefont {Nori},\ and\ \citenamefont {Plenio}}]{Burgarth2010}%
  \BibitemOpen
  \bibfield  {author} {\bibinfo {author} {\bibfnamefont {D.}~\bibnamefont {Burgarth}}, \bibinfo {author} {\bibfnamefont {K.}~\bibnamefont {Maruyama}}, \bibinfo {author} {\bibfnamefont {M.}~\bibnamefont {Murphy}}, \bibinfo {author} {\bibfnamefont {S.}~\bibnamefont {Montangero}}, \bibinfo {author} {\bibfnamefont {T.}~\bibnamefont {Calarco}}, \bibinfo {author} {\bibfnamefont {F.}~\bibnamefont {Nori}}, \ and\ \bibinfo {author} {\bibfnamefont {M.~B.}\ \bibnamefont {Plenio}},\ }\href {\doibase 10.1103/PhysRevA.81.040303} {\bibfield  {journal} {\bibinfo  {journal} {Phys. Rev. A}\ }\textbf {\bibinfo {volume} {81}},\ \bibinfo {pages} {040303} (\bibinfo {year} {2010})}\BibitemShut {NoStop}%
\bibitem [{\citenamefont {Yang}\ \emph {et~al.}(2010)\citenamefont {Yang}, \citenamefont {Bayat},\ and\ \citenamefont {Bose}}]{Yang2010}%
  \BibitemOpen
  \bibfield  {author} {\bibinfo {author} {\bibfnamefont {S.}~\bibnamefont {Yang}}, \bibinfo {author} {\bibfnamefont {A.}~\bibnamefont {Bayat}}, \ and\ \bibinfo {author} {\bibfnamefont {S.}~\bibnamefont {Bose}},\ }\href {\doibase 10.1103/PhysRevA.82.022336} {\bibfield  {journal} {\bibinfo  {journal} {Phys. Rev. A}\ }\textbf {\bibinfo {volume} {82}},\ \bibinfo {pages} {022336} (\bibinfo {year} {2010})}\BibitemShut {NoStop}%
\bibitem [{\citenamefont {Gong}\ and\ \citenamefont {Brumer}(2007)}]{Gong2007}%
  \BibitemOpen
  \bibfield  {author} {\bibinfo {author} {\bibfnamefont {J.}~\bibnamefont {Gong}}\ and\ \bibinfo {author} {\bibfnamefont {P.}~\bibnamefont {Brumer}},\ }\href {\doibase 10.1103/PhysRevA.75.032331} {\bibfield  {journal} {\bibinfo  {journal} {Phys. Rev. A}\ }\textbf {\bibinfo {volume} {75}},\ \bibinfo {pages} {032331} (\bibinfo {year} {2007})}\BibitemShut {NoStop}%
\bibitem [{\citenamefont {Murphy}\ \emph {et~al.}(2010)\citenamefont {Murphy}, \citenamefont {Montangero}, \citenamefont {Giovannetti},\ and\ \citenamefont {Calarco}}]{Murphy2010}%
  \BibitemOpen
  \bibfield  {author} {\bibinfo {author} {\bibfnamefont {M.}~\bibnamefont {Murphy}}, \bibinfo {author} {\bibfnamefont {S.}~\bibnamefont {Montangero}}, \bibinfo {author} {\bibfnamefont {V.}~\bibnamefont {Giovannetti}}, \ and\ \bibinfo {author} {\bibfnamefont {T.}~\bibnamefont {Calarco}},\ }\href {\doibase 10.1103/PhysRevA.82.022318} {\bibfield  {journal} {\bibinfo  {journal} {Phys. Rev. A}\ }\textbf {\bibinfo {volume} {82}},\ \bibinfo {pages} {022318} (\bibinfo {year} {2010})}\BibitemShut {NoStop}%
\bibitem [{\citenamefont {Lemonde}\ \emph {et~al.}(2019)\citenamefont {Lemonde}, \citenamefont {Peano}, \citenamefont {Rabl},\ and\ \citenamefont {Angelakis}}]{Lemonde2019}%
  \BibitemOpen
  \bibfield  {author} {\bibinfo {author} {\bibfnamefont {M.-A.}\ \bibnamefont {Lemonde}}, \bibinfo {author} {\bibfnamefont {V.}~\bibnamefont {Peano}}, \bibinfo {author} {\bibfnamefont {P.}~\bibnamefont {Rabl}}, \ and\ \bibinfo {author} {\bibfnamefont {D.~G.}\ \bibnamefont {Angelakis}},\ }\href {\doibase 10.1088/1367-2630/ab51f5} {\bibfield  {journal} {\bibinfo  {journal} {New Journal of Physics}\ }\textbf {\bibinfo {volume} {21}},\ \bibinfo {pages} {113030} (\bibinfo {year} {2019})}\BibitemShut {NoStop}%
\bibitem [{\citenamefont {Romero}\ \emph {et~al.}(2024)\citenamefont {Romero}, \citenamefont {Chen}, \citenamefont {Platero},\ and\ \citenamefont {Ban}}]{Romero2024}%
  \BibitemOpen
  \bibfield  {author} {\bibinfo {author} {\bibfnamefont {S.~V.}\ \bibnamefont {Romero}}, \bibinfo {author} {\bibfnamefont {X.}~\bibnamefont {Chen}}, \bibinfo {author} {\bibfnamefont {G.}~\bibnamefont {Platero}}, \ and\ \bibinfo {author} {\bibfnamefont {Y.}~\bibnamefont {Ban}},\ }\href {\doibase 10.1103/PhysRevApplied.21.034033} {\bibfield  {journal} {\bibinfo  {journal} {Phys. Rev. Appl.}\ }\textbf {\bibinfo {volume} {21}},\ \bibinfo {pages} {034033} (\bibinfo {year} {2024})}\BibitemShut {NoStop}%
\bibitem [{\citenamefont {Wang}\ \emph {et~al.}(2024)\citenamefont {Wang}, \citenamefont {Fan}, \citenamefont {Chen}, \citenamefont {Jiang}, \citenamefont {Gao}, \citenamefont {Lado},\ and\ \citenamefont {Yang}}]{wang2024TopoMagneticConstruction}%
  \BibitemOpen
  \bibfield  {author} {\bibinfo {author} {\bibfnamefont {H.}~\bibnamefont {Wang}}, \bibinfo {author} {\bibfnamefont {P.}~\bibnamefont {Fan}}, \bibinfo {author} {\bibfnamefont {J.}~\bibnamefont {Chen}}, \bibinfo {author} {\bibfnamefont {L.}~\bibnamefont {Jiang}}, \bibinfo {author} {\bibfnamefont {H.-J.}\ \bibnamefont {Gao}}, \bibinfo {author} {\bibfnamefont {J.~L.}\ \bibnamefont {Lado}}, \ and\ \bibinfo {author} {\bibfnamefont {K.}~\bibnamefont {Yang}},\ }\href@noop {} {\bibfield  {journal} {\bibinfo  {journal} {Nature Nanotechnology}\ }\textbf {\bibinfo {volume} {19}},\ \bibinfo {pages} {1782} (\bibinfo {year} {2024})}\BibitemShut {NoStop}%
\bibitem [{\citenamefont {Baumann}\ \emph {et~al.}(2015)\citenamefont {Baumann}, \citenamefont {Donati}, \citenamefont {Stepanow}, \citenamefont {Rusponi}, \citenamefont {Paul}, \citenamefont {Gangopadhyay}, \citenamefont {Rau}, \citenamefont {Pacchioni}, \citenamefont {Gragnaniello}, \citenamefont {Pivetta} \emph {et~al.}}]{baumann2015b}%
  \BibitemOpen
  \bibfield  {author} {\bibinfo {author} {\bibfnamefont {S.}~\bibnamefont {Baumann}}, \bibinfo {author} {\bibfnamefont {F.}~\bibnamefont {Donati}}, \bibinfo {author} {\bibfnamefont {S.}~\bibnamefont {Stepanow}}, \bibinfo {author} {\bibfnamefont {S.}~\bibnamefont {Rusponi}}, \bibinfo {author} {\bibfnamefont {W.}~\bibnamefont {Paul}}, \bibinfo {author} {\bibfnamefont {S.}~\bibnamefont {Gangopadhyay}}, \bibinfo {author} {\bibfnamefont {I.}~\bibnamefont {Rau}}, \bibinfo {author} {\bibfnamefont {G.}~\bibnamefont {Pacchioni}}, \bibinfo {author} {\bibfnamefont {L.}~\bibnamefont {Gragnaniello}}, \bibinfo {author} {\bibfnamefont {M.}~\bibnamefont {Pivetta}},  \emph {et~al.},\ }\href@noop {} {\bibfield  {journal} {\bibinfo  {journal} {Physical review letters}\ }\textbf {\bibinfo {volume} {115}},\ \bibinfo {pages} {237202} (\bibinfo {year} {2015})}\BibitemShut {NoStop}%
\bibitem [{\citenamefont {Zhao}\ \emph {et~al.}(2024)\citenamefont {Zhao}, \citenamefont {Catarina}, \citenamefont {Zhang}, \citenamefont {Henriques}, \citenamefont {Yang}, \citenamefont {Ma}, \citenamefont {Feng}, \citenamefont {Gr{\"o}ning}, \citenamefont {Ruffieux}, \citenamefont {Fern{\'a}ndez-Rossier} \emph {et~al.}}]{zhao2024nanografeno1}%
  \BibitemOpen
  \bibfield  {author} {\bibinfo {author} {\bibfnamefont {C.}~\bibnamefont {Zhao}}, \bibinfo {author} {\bibfnamefont {G.}~\bibnamefont {Catarina}}, \bibinfo {author} {\bibfnamefont {J.-J.}\ \bibnamefont {Zhang}}, \bibinfo {author} {\bibfnamefont {J.~C.}\ \bibnamefont {Henriques}}, \bibinfo {author} {\bibfnamefont {L.}~\bibnamefont {Yang}}, \bibinfo {author} {\bibfnamefont {J.}~\bibnamefont {Ma}}, \bibinfo {author} {\bibfnamefont {X.}~\bibnamefont {Feng}}, \bibinfo {author} {\bibfnamefont {O.}~\bibnamefont {Gr{\"o}ning}}, \bibinfo {author} {\bibfnamefont {P.}~\bibnamefont {Ruffieux}}, \bibinfo {author} {\bibfnamefont {J.}~\bibnamefont {Fern{\'a}ndez-Rossier}},  \emph {et~al.},\ }\href@noop {} {\bibfield  {journal} {\bibinfo  {journal} {Nature Nanotechnology}\ }\textbf {\bibinfo {volume} {19}},\ \bibinfo {pages} {1789} (\bibinfo {year} {2024})}\BibitemShut {NoStop}%
\bibitem [{\citenamefont {Zhao}\ \emph {et~al.}(2025)\citenamefont {Zhao}, \citenamefont {Yang}, \citenamefont {Henriques}, \citenamefont {Ferri-Cort{\'e}s}, \citenamefont {Catarina}, \citenamefont {Pignedoli}, \citenamefont {Ma}, \citenamefont {Feng}, \citenamefont {Ruffieux}, \citenamefont {Fern{\'a}ndez-Rossier} \emph {et~al.}}]{zhao2025nanografeno2}%
  \BibitemOpen
  \bibfield  {author} {\bibinfo {author} {\bibfnamefont {C.}~\bibnamefont {Zhao}}, \bibinfo {author} {\bibfnamefont {L.}~\bibnamefont {Yang}}, \bibinfo {author} {\bibfnamefont {J.~C.}\ \bibnamefont {Henriques}}, \bibinfo {author} {\bibfnamefont {M.}~\bibnamefont {Ferri-Cort{\'e}s}}, \bibinfo {author} {\bibfnamefont {G.}~\bibnamefont {Catarina}}, \bibinfo {author} {\bibfnamefont {C.~A.}\ \bibnamefont {Pignedoli}}, \bibinfo {author} {\bibfnamefont {J.}~\bibnamefont {Ma}}, \bibinfo {author} {\bibfnamefont {X.}~\bibnamefont {Feng}}, \bibinfo {author} {\bibfnamefont {P.}~\bibnamefont {Ruffieux}}, \bibinfo {author} {\bibfnamefont {J.}~\bibnamefont {Fern{\'a}ndez-Rossier}},  \emph {et~al.},\ }\href@noop {} {\bibfield  {journal} {\bibinfo  {journal} {Nature Materials}\ ,\ \bibinfo {pages} {1}} (\bibinfo {year} {2025})}\BibitemShut {NoStop}%
\bibitem [{\citenamefont {Fu}\ \emph {et~al.}(2025)\citenamefont {Fu}, \citenamefont {Huang}, \citenamefont {Liu}, \citenamefont {Henriques}, \citenamefont {Gao}, \citenamefont {Han}, \citenamefont {Chen}, \citenamefont {Wang}, \citenamefont {Palma}, \citenamefont {Cheng} \emph {et~al.}}]{fu2025nanografenobuilding}%
  \BibitemOpen
  \bibfield  {author} {\bibinfo {author} {\bibfnamefont {X.}~\bibnamefont {Fu}}, \bibinfo {author} {\bibfnamefont {L.}~\bibnamefont {Huang}}, \bibinfo {author} {\bibfnamefont {K.}~\bibnamefont {Liu}}, \bibinfo {author} {\bibfnamefont {J.~C.}\ \bibnamefont {Henriques}}, \bibinfo {author} {\bibfnamefont {Y.}~\bibnamefont {Gao}}, \bibinfo {author} {\bibfnamefont {X.}~\bibnamefont {Han}}, \bibinfo {author} {\bibfnamefont {H.}~\bibnamefont {Chen}}, \bibinfo {author} {\bibfnamefont {Y.}~\bibnamefont {Wang}}, \bibinfo {author} {\bibfnamefont {C.-A.}\ \bibnamefont {Palma}}, \bibinfo {author} {\bibfnamefont {Z.}~\bibnamefont {Cheng}},  \emph {et~al.},\ }\href@noop {} {\bibfield  {journal} {\bibinfo  {journal} {Nature Synthesis}\ ,\ \bibinfo {pages} {1}} (\bibinfo {year} {2025})}\BibitemShut {NoStop}%
\bibitem [{\citenamefont {Li}\ \emph {et~al.}(2019)\citenamefont {Li}, \citenamefont {Zhang},\ and\ \citenamefont {Lin}}]{Li2018TopoPhase}%
  \BibitemOpen
  \bibfield  {author} {\bibinfo {author} {\bibfnamefont {Y.-C.}\ \bibnamefont {Li}}, \bibinfo {author} {\bibfnamefont {J.}~\bibnamefont {Zhang}}, \ and\ \bibinfo {author} {\bibfnamefont {H.-Q.}\ \bibnamefont {Lin}},\ }\href@noop {} {\bibfield  {journal} {\bibinfo  {journal} {Physical Review B}\ }\textbf {\bibinfo {volume} {99}},\ \bibinfo {pages} {205424} (\bibinfo {year} {2019})}\BibitemShut {NoStop}%
\bibitem [{\citenamefont {Apollaro}\ \emph {et~al.}(2022)\citenamefont {Apollaro}, \citenamefont {Lorenzo}, \citenamefont {Plastina}, \citenamefont {Consiglio},\ and\ \citenamefont {Życzkowski}}]{Apollaro2022}%
  \BibitemOpen
  \bibfield  {author} {\bibinfo {author} {\bibfnamefont {T.~J.~G.}\ \bibnamefont {Apollaro}}, \bibinfo {author} {\bibfnamefont {S.}~\bibnamefont {Lorenzo}}, \bibinfo {author} {\bibfnamefont {F.}~\bibnamefont {Plastina}}, \bibinfo {author} {\bibfnamefont {M.}~\bibnamefont {Consiglio}}, \ and\ \bibinfo {author} {\bibfnamefont {K.}~\bibnamefont {Życzkowski}},\ }\href {\doibase 10.1088/1367-2630/ac86e7} {\bibfield  {journal} {\bibinfo  {journal} {New Journal of Physics}\ }\textbf {\bibinfo {volume} {24}},\ \bibinfo {pages} {083025} (\bibinfo {year} {2022})}\BibitemShut {NoStop}%
\bibitem [{\citenamefont {Serra}\ \emph {et~al.}(2022{\natexlab{a}})\citenamefont {Serra}, \citenamefont {Ferrón},\ and\ \citenamefont {Osenda}}]{Serra2022PLA}%
  \BibitemOpen
  \bibfield  {author} {\bibinfo {author} {\bibfnamefont {P.}~\bibnamefont {Serra}}, \bibinfo {author} {\bibfnamefont {A.}~\bibnamefont {Ferrón}}, \ and\ \bibinfo {author} {\bibfnamefont {O.}~\bibnamefont {Osenda}},\ }\href {\doibase https://doi.org/10.1016/j.physleta.2022.128362} {\bibfield  {journal} {\bibinfo  {journal} {Physics Letters A}\ }\textbf {\bibinfo {volume} {449}},\ \bibinfo {pages} {128362} (\bibinfo {year} {2022}{\natexlab{a}})}\BibitemShut {NoStop}%
\bibitem [{\citenamefont {Ferr{\'o}n}\ \emph {et~al.}(2022)\citenamefont {Ferr{\'o}n}, \citenamefont {Serra},\ and\ \citenamefont {Osenda}}]{Ferron2021UnderstandingTP}%
  \BibitemOpen
  \bibfield  {author} {\bibinfo {author} {\bibfnamefont {A.}~\bibnamefont {Ferr{\'o}n}}, \bibinfo {author} {\bibfnamefont {P.}~\bibnamefont {Serra}}, \ and\ \bibinfo {author} {\bibfnamefont {O.}~\bibnamefont {Osenda}},\ }\href@noop {} {\bibfield  {journal} {\bibinfo  {journal} {Physica Scripta}\ }\textbf {\bibinfo {volume} {97}},\ \bibinfo {pages} {115103} (\bibinfo {year} {2022})}\BibitemShut {NoStop}%
\bibitem [{\citenamefont {Serra}\ \emph {et~al.}(2022{\natexlab{b}})\citenamefont {Serra}, \citenamefont {Ferrón},\ and\ \citenamefont {Osenda}}]{Serra2022JPA}%
  \BibitemOpen
  \bibfield  {author} {\bibinfo {author} {\bibfnamefont {P.}~\bibnamefont {Serra}}, \bibinfo {author} {\bibfnamefont {A.}~\bibnamefont {Ferrón}}, \ and\ \bibinfo {author} {\bibfnamefont {O.}~\bibnamefont {Osenda}},\ }\href {\doibase 10.1088/1751-8121/ac901d} {\bibfield  {journal} {\bibinfo  {journal} {Journal of Physics A: Mathematical and Theoretical}\ }\textbf {\bibinfo {volume} {55}},\ \bibinfo {pages} {405302} (\bibinfo {year} {2022}{\natexlab{b}})}\BibitemShut {NoStop}%
\bibitem [{\citenamefont {Werschnik}\ and\ \citenamefont {Gross}(2007)}]{Werschnik2007}%
  \BibitemOpen
  \bibfield  {author} {\bibinfo {author} {\bibfnamefont {J.}~\bibnamefont {Werschnik}}\ and\ \bibinfo {author} {\bibfnamefont {E.~K.~U.}\ \bibnamefont {Gross}},\ }\href {\doibase 10.1088/0953-4075/40/18/R01} {\bibfield  {journal} {\bibinfo  {journal} {Journal of Physics B: Atomic, Molecular and Optical Physics}\ }\textbf {\bibinfo {volume} {40}},\ \bibinfo {pages} {R175} (\bibinfo {year} {2007})}\BibitemShut {NoStop}%
\bibitem [{\citenamefont {Zhang}\ \emph {et~al.}(2018)\citenamefont {Zhang}, \citenamefont {Cui}, \citenamefont {Wang},\ and\ \citenamefont {Yung}}]{Zhang2018}%
  \BibitemOpen
  \bibfield  {author} {\bibinfo {author} {\bibfnamefont {X.-M.}\ \bibnamefont {Zhang}}, \bibinfo {author} {\bibfnamefont {Z.-W.}\ \bibnamefont {Cui}}, \bibinfo {author} {\bibfnamefont {X.}~\bibnamefont {Wang}}, \ and\ \bibinfo {author} {\bibfnamefont {M.-H.}\ \bibnamefont {Yung}},\ }\href {\doibase 10.1103/PhysRevA.97.052333} {\bibfield  {journal} {\bibinfo  {journal} {Phys. Rev. A}\ }\textbf {\bibinfo {volume} {97}},\ \bibinfo {pages} {052333} (\bibinfo {year} {2018})}\BibitemShut {NoStop}%
\bibitem [{\citenamefont {Bukov}\ \emph {et~al.}(2018)\citenamefont {Bukov}, \citenamefont {Day}, \citenamefont {Sels}, \citenamefont {Weinberg}, \citenamefont {Polkovnikov},\ and\ \citenamefont {Mehta}}]{Bukov2018}%
  \BibitemOpen
  \bibfield  {author} {\bibinfo {author} {\bibfnamefont {M.}~\bibnamefont {Bukov}}, \bibinfo {author} {\bibfnamefont {A.~G.~R.}\ \bibnamefont {Day}}, \bibinfo {author} {\bibfnamefont {D.}~\bibnamefont {Sels}}, \bibinfo {author} {\bibfnamefont {P.}~\bibnamefont {Weinberg}}, \bibinfo {author} {\bibfnamefont {A.}~\bibnamefont {Polkovnikov}}, \ and\ \bibinfo {author} {\bibfnamefont {P.}~\bibnamefont {Mehta}},\ }\href {\doibase 10.1103/PhysRevX.8.031086} {\bibfield  {journal} {\bibinfo  {journal} {Phys. Rev. X}\ }\textbf {\bibinfo {volume} {8}},\ \bibinfo {pages} {031086} (\bibinfo {year} {2018})}\BibitemShut {NoStop}%
\bibitem [{\citenamefont {Sivak}\ \emph {et~al.}(2022)\citenamefont {Sivak}, \citenamefont {Eickbusch}, \citenamefont {Liu}, \citenamefont {Royer}, \citenamefont {Tsioutsios},\ and\ \citenamefont {Devoret}}]{Sivak2022}%
  \BibitemOpen
  \bibfield  {author} {\bibinfo {author} {\bibfnamefont {V.~V.}\ \bibnamefont {Sivak}}, \bibinfo {author} {\bibfnamefont {A.}~\bibnamefont {Eickbusch}}, \bibinfo {author} {\bibfnamefont {H.}~\bibnamefont {Liu}}, \bibinfo {author} {\bibfnamefont {B.}~\bibnamefont {Royer}}, \bibinfo {author} {\bibfnamefont {I.}~\bibnamefont {Tsioutsios}}, \ and\ \bibinfo {author} {\bibfnamefont {M.~H.}\ \bibnamefont {Devoret}},\ }\href {\doibase 10.1103/PhysRevX.12.011059} {\bibfield  {journal} {\bibinfo  {journal} {Phys. Rev. X}\ }\textbf {\bibinfo {volume} {12}},\ \bibinfo {pages} {011059} (\bibinfo {year} {2022})}\BibitemShut {NoStop}%
\bibitem [{\citenamefont {Porotti}\ \emph {et~al.}(2022)\citenamefont {Porotti}, \citenamefont {Essig}, \citenamefont {Huard},\ and\ \citenamefont {Marquardt}}]{Porotti2022}%
  \BibitemOpen
  \bibfield  {author} {\bibinfo {author} {\bibfnamefont {R.}~\bibnamefont {Porotti}}, \bibinfo {author} {\bibfnamefont {A.}~\bibnamefont {Essig}}, \bibinfo {author} {\bibfnamefont {B.}~\bibnamefont {Huard}}, \ and\ \bibinfo {author} {\bibfnamefont {F.}~\bibnamefont {Marquardt}},\ }\href {\doibase 10.22331/q-2022-06-28-747} {\bibfield  {journal} {\bibinfo  {journal} {{Quantum}}\ }\textbf {\bibinfo {volume} {6}},\ \bibinfo {pages} {747} (\bibinfo {year} {2022})}\BibitemShut {NoStop}%
\bibitem [{\citenamefont {Niu}\ \emph {et~al.}(2019)\citenamefont {Niu}, \citenamefont {Boixo}, \citenamefont {Smelyanskiy},\ and\ \citenamefont {Neven}}]{Niu2019}%
  \BibitemOpen
  \bibfield  {author} {\bibinfo {author} {\bibfnamefont {M.~Y.}\ \bibnamefont {Niu}}, \bibinfo {author} {\bibfnamefont {S.}~\bibnamefont {Boixo}}, \bibinfo {author} {\bibfnamefont {V.~N.}\ \bibnamefont {Smelyanskiy}}, \ and\ \bibinfo {author} {\bibfnamefont {H.}~\bibnamefont {Neven}},\ }\href@noop {} {\bibfield  {journal} {\bibinfo  {journal} {npj Quantum Information}\ }\textbf {\bibinfo {volume} {5}},\ \bibinfo {pages} {33} (\bibinfo {year} {2019})}\BibitemShut {NoStop}%
\bibitem [{\citenamefont {Acosta~Coden}\ \emph {et~al.}(2024)\citenamefont {Acosta~Coden}, \citenamefont {Osenda},\ and\ \citenamefont {Ferrón}}]{coden2024quantum}%
  \BibitemOpen
  \bibfield  {author} {\bibinfo {author} {\bibfnamefont {D.~S.}\ \bibnamefont {Acosta~Coden}}, \bibinfo {author} {\bibfnamefont {O.}~\bibnamefont {Osenda}}, \ and\ \bibinfo {author} {\bibfnamefont {A.}~\bibnamefont {Ferrón}},\ }\href {\doibase 10.1088/1361-6455/ad9a30} {\bibfield  {journal} {\bibinfo  {journal} {Journal of Physics B: Atomic, Molecular and Optical Physics}\ }\textbf {\bibinfo {volume} {58}},\ \bibinfo {pages} {015504} (\bibinfo {year} {2024})}\BibitemShut {NoStop}%
\bibitem [{\citenamefont {Zwick}\ \emph {et~al.}(2015)\citenamefont {Zwick}, \citenamefont {Alvarez}, \citenamefont {Stolze},\ and\ \citenamefont {Osenda}}]{Zwick2015}%
  \BibitemOpen
  \bibfield  {author} {\bibinfo {author} {\bibfnamefont {A.}~\bibnamefont {Zwick}}, \bibinfo {author} {\bibfnamefont {G.~A.}\ \bibnamefont {Alvarez}}, \bibinfo {author} {\bibfnamefont {J.}~\bibnamefont {Stolze}}, \ and\ \bibinfo {author} {\bibfnamefont {O.}~\bibnamefont {Osenda}},\ }\href {https://www.rintonpress.com/xxqic15/qic-15-78/0582-0600.pdf} {\bibfield  {journal} {\bibinfo  {journal} {Quantum Information \& Computation}\ }\textbf {\bibinfo {volume} {15}},\ \bibinfo {pages} {0582} (\bibinfo {year} {2015})}\BibitemShut {NoStop}%
\bibitem [{\citenamefont {Stolze}\ \emph {et~al.}(2014)\citenamefont {Stolze}, \citenamefont {{\'A}lvarez}, \citenamefont {Osenda},\ and\ \citenamefont {Zwick}}]{Stolze2014}%
  \BibitemOpen
  \bibfield  {author} {\bibinfo {author} {\bibfnamefont {J.}~\bibnamefont {Stolze}}, \bibinfo {author} {\bibfnamefont {G.~A.}\ \bibnamefont {{\'A}lvarez}}, \bibinfo {author} {\bibfnamefont {O.}~\bibnamefont {Osenda}}, \ and\ \bibinfo {author} {\bibfnamefont {A.}~\bibnamefont {Zwick}},\ }\enquote {\bibinfo {title} {Robustness of spin-chain state-transfer schemes},}\ in\ \href {\doibase 10.1007/978-3-642-39937-4_5} {\emph {\bibinfo {booktitle} {Quantum State Transfer and Network Engineering}}},\ \bibinfo {editor} {edited by\ \bibinfo {editor} {\bibfnamefont {G.~M.}\ \bibnamefont {Nikolopoulos}}\ and\ \bibinfo {editor} {\bibfnamefont {I.}~\bibnamefont {Jex}}}\ (\bibinfo  {publisher} {Springer Berlin Heidelberg},\ \bibinfo {address} {Berlin, Heidelberg},\ \bibinfo {year} {2014})\ pp.\ \bibinfo {pages} {149--182}\BibitemShut {NoStop}%
\bibitem [{\citenamefont {Petrosyan}\ \emph {et~al.}(2010)\citenamefont {Petrosyan}, \citenamefont {Nikolopoulos},\ and\ \citenamefont {Lambropoulos}}]{Petrosyan2010}%
  \BibitemOpen
  \bibfield  {author} {\bibinfo {author} {\bibfnamefont {D.}~\bibnamefont {Petrosyan}}, \bibinfo {author} {\bibfnamefont {G.~M.}\ \bibnamefont {Nikolopoulos}}, \ and\ \bibinfo {author} {\bibfnamefont {P.}~\bibnamefont {Lambropoulos}},\ }\href {\doibase 10.1103/PhysRevA.81.042307} {\bibfield  {journal} {\bibinfo  {journal} {Phys. Rev. A}\ }\textbf {\bibinfo {volume} {81}},\ \bibinfo {pages} {042307} (\bibinfo {year} {2010})}\BibitemShut {NoStop}%
\bibitem [{\citenamefont {Bose}(2003)}]{bose2003quantum}%
  \BibitemOpen
  \bibfield  {author} {\bibinfo {author} {\bibfnamefont {S.}~\bibnamefont {Bose}},\ }\href {\doibase 10.1103/PhysRevLett.91.207901} {\bibfield  {journal} {\bibinfo  {journal} {Phys. Rev. Lett.}\ }\textbf {\bibinfo {volume} {91}},\ \bibinfo {pages} {207901} (\bibinfo {year} {2003})}\BibitemShut {NoStop}%
\bibitem [{\citenamefont {Serra}\ \emph {et~al.}(2023)\citenamefont {Serra}, \citenamefont {Ferrón},\ and\ \citenamefont {Osenda}}]{serra2023scaling}%
  \BibitemOpen
  \bibfield  {author} {\bibinfo {author} {\bibfnamefont {P.}~\bibnamefont {Serra}}, \bibinfo {author} {\bibfnamefont {A.}~\bibnamefont {Ferrón}}, \ and\ \bibinfo {author} {\bibfnamefont {O.}~\bibnamefont {Osenda}},\ }\href {\doibase 10.1088/1751-8121/ad0d20} {\bibfield  {journal} {\bibinfo  {journal} {Journal of Physics A: Mathematical and Theoretical}\ }\textbf {\bibinfo {volume} {57}},\ \bibinfo {pages} {015304} (\bibinfo {year} {2023})}\BibitemShut {NoStop}%
\bibitem [{\citenamefont {Kay}(2022)}]{kay2022PTheisen}%
  \BibitemOpen
  \bibfield  {author} {\bibinfo {author} {\bibfnamefont {A.}~\bibnamefont {Kay}},\ }\href {https://arxiv.org/abs/1906.06223} {\enquote {\bibinfo {title} {The limits of quantum state transfer for field-free heisenberg chains},}\ } (\bibinfo {year} {2022}),\ \Eprint {http://arxiv.org/abs/1906.06223} {arXiv:1906.06223 [quant-ph]} \BibitemShut {NoStop}%
\bibitem [{\citenamefont {Peirce}\ \emph {et~al.}(1988)\citenamefont {Peirce}, \citenamefont {Dahleh},\ and\ \citenamefont {Rabitz}}]{Peirce1988OCT}%
  \BibitemOpen
  \bibfield  {author} {\bibinfo {author} {\bibfnamefont {A.~P.}\ \bibnamefont {Peirce}}, \bibinfo {author} {\bibfnamefont {M.~A.}\ \bibnamefont {Dahleh}}, \ and\ \bibinfo {author} {\bibfnamefont {H.}~\bibnamefont {Rabitz}},\ }\href {\doibase 10.1103/PhysRevA.37.4950} {\bibfield  {journal} {\bibinfo  {journal} {Phys. Rev. A}\ }\textbf {\bibinfo {volume} {37}},\ \bibinfo {pages} {4950} (\bibinfo {year} {1988})}\BibitemShut {NoStop}%
\bibitem [{\citenamefont {Dong}\ and\ \citenamefont {Petersen}(2010)}]{Dong2016OCT}%
  \BibitemOpen
  \bibfield  {author} {\bibinfo {author} {\bibfnamefont {D.}~\bibnamefont {Dong}}\ and\ \bibinfo {author} {\bibfnamefont {I.}~\bibnamefont {Petersen}},\ }\href {\doibase 10.1049/iet-cta.2009.0508} {\bibfield  {journal} {\bibinfo  {journal} {IET Control Theory \& Applications}\ }\textbf {\bibinfo {volume} {4}},\ \bibinfo {pages} {2651} (\bibinfo {year} {2010})},\ \Eprint {http://arxiv.org/abs/https://digital-library.theiet.org/doi/pdf/10.1049/iet-cta.2009.0508} {https://digital-library.theiet.org/doi/pdf/10.1049/iet-cta.2009.0508} \BibitemShut {NoStop}%
\bibitem [{\citenamefont {Hull}(2010)}]{Hull2010OCT}%
  \BibitemOpen
  \bibfield  {author} {\bibinfo {author} {\bibfnamefont {D.~G.}\ \bibnamefont {Hull}},\ }\href@noop {} {\emph {\bibinfo {title} {Optimal Control Theory for Applications}}},\ Mechanical Engineering Series\ (\bibinfo  {publisher} {Springer},\ \bibinfo {address} {New York, NY},\ \bibinfo {year} {2010})\BibitemShut {NoStop}%
\bibitem [{\citenamefont {James}(2021)}]{James2021OCT}%
  \BibitemOpen
  \bibfield  {author} {\bibinfo {author} {\bibfnamefont {M.~R.}\ \bibnamefont {James}},\ }\href@noop {} {\bibfield  {journal} {\bibinfo  {journal} {Annu. Rev. Control Robot. Auton. Syst.}\ }\textbf {\bibinfo {volume} {4}},\ \bibinfo {pages} {343} (\bibinfo {year} {2021})}\BibitemShut {NoStop}%
\bibitem [{\citenamefont {Yang}\ \emph {et~al.}(2017)\citenamefont {Yang}, \citenamefont {Bae}, \citenamefont {Paul}, \citenamefont {Natterer}, \citenamefont {Willke}, \citenamefont {Lado}, \citenamefont {Ferr{\'o}n}, \citenamefont {Choi}, \citenamefont {Fern{\'a}ndez-Rossier}, \citenamefont {Heinrich} \emph {et~al.}}]{yang2017engineering}%
  \BibitemOpen
  \bibfield  {author} {\bibinfo {author} {\bibfnamefont {K.}~\bibnamefont {Yang}}, \bibinfo {author} {\bibfnamefont {Y.}~\bibnamefont {Bae}}, \bibinfo {author} {\bibfnamefont {W.}~\bibnamefont {Paul}}, \bibinfo {author} {\bibfnamefont {F.~D.}\ \bibnamefont {Natterer}}, \bibinfo {author} {\bibfnamefont {P.}~\bibnamefont {Willke}}, \bibinfo {author} {\bibfnamefont {J.~L.}\ \bibnamefont {Lado}}, \bibinfo {author} {\bibfnamefont {A.}~\bibnamefont {Ferr{\'o}n}}, \bibinfo {author} {\bibfnamefont {T.}~\bibnamefont {Choi}}, \bibinfo {author} {\bibfnamefont {J.}~\bibnamefont {Fern{\'a}ndez-Rossier}}, \bibinfo {author} {\bibfnamefont {A.~J.}\ \bibnamefont {Heinrich}},  \emph {et~al.},\ }\href@noop {} {\bibfield  {journal} {\bibinfo  {journal} {Physical review letters}\ }\textbf {\bibinfo {volume} {119}},\ \bibinfo {pages} {227206} (\bibinfo {year} {2017})}\BibitemShut {NoStop}%
\bibitem [{\citenamefont {Wang}\ \emph {et~al.}(2025)\citenamefont {Wang}, \citenamefont {Chen}, \citenamefont {Fan}, \citenamefont {del Castillo}, \citenamefont {Ferr{\'o}n}, \citenamefont {Jiang}, \citenamefont {Wu}, \citenamefont {Li}, \citenamefont {Gao}, \citenamefont {Fan} \emph {et~al.}}]{wang2025electrically}%
  \BibitemOpen
  \bibfield  {author} {\bibinfo {author} {\bibfnamefont {H.}~\bibnamefont {Wang}}, \bibinfo {author} {\bibfnamefont {J.}~\bibnamefont {Chen}}, \bibinfo {author} {\bibfnamefont {P.}~\bibnamefont {Fan}}, \bibinfo {author} {\bibfnamefont {Y.}~\bibnamefont {del Castillo}}, \bibinfo {author} {\bibfnamefont {A.}~\bibnamefont {Ferr{\'o}n}}, \bibinfo {author} {\bibfnamefont {L.}~\bibnamefont {Jiang}}, \bibinfo {author} {\bibfnamefont {Z.}~\bibnamefont {Wu}}, \bibinfo {author} {\bibfnamefont {S.}~\bibnamefont {Li}}, \bibinfo {author} {\bibfnamefont {H.-J.}\ \bibnamefont {Gao}}, \bibinfo {author} {\bibfnamefont {H.}~\bibnamefont {Fan}},  \emph {et~al.},\ }\href@noop {} {\bibfield  {journal} {\bibinfo  {journal} {arXiv preprint arXiv:2506.01033}\ } (\bibinfo {year} {2025})}\BibitemShut {NoStop}%
\bibitem [{\citenamefont {Kot}\ \emph {et~al.}(2023)\citenamefont {Kot}, \citenamefont {Ismail}, \citenamefont {Drost}, \citenamefont {Siebrecht}, \citenamefont {Huang},\ and\ \citenamefont {Ast}}]{kot2023electric}%
  \BibitemOpen
  \bibfield  {author} {\bibinfo {author} {\bibfnamefont {P.}~\bibnamefont {Kot}}, \bibinfo {author} {\bibfnamefont {M.}~\bibnamefont {Ismail}}, \bibinfo {author} {\bibfnamefont {R.}~\bibnamefont {Drost}}, \bibinfo {author} {\bibfnamefont {J.}~\bibnamefont {Siebrecht}}, \bibinfo {author} {\bibfnamefont {H.}~\bibnamefont {Huang}}, \ and\ \bibinfo {author} {\bibfnamefont {C.~R.}\ \bibnamefont {Ast}},\ }\href@noop {} {\bibfield  {journal} {\bibinfo  {journal} {Nature Communications}\ }\textbf {\bibinfo {volume} {14}},\ \bibinfo {pages} {6612} (\bibinfo {year} {2023})}\BibitemShut {NoStop}%
\end{thebibliography}%

\end{document}